\documentclass[9pt,lineno]{elife}

\usepackage[utf8]{inputenc}
\usepackage{import}
\usepackage{array}
\usepackage{tabularx}
\usepackage{wrapfig}
\usepackage{textcomp}
\usepackage{amsmath}
\usepackage{graphicx}
\usepackage{esint}
\usepackage{subscript}
\usepackage{rotating}
\usepackage{float}
\usepackage{graphicx,kantlipsum,setspace}
\usepackage{tikz}
\usepackage{tikzscale}
\usepackage{pgfplots}
\usepackage{lmodern}

\usepackage{caption}
\usetikzlibrary{arrows}
\makeatletter
\def\ps@pprintTitle{%
 \let\@oddhead\@empty
 \let\@evenhead\@empty
 \def\@oddfoot{\centerline{\thepage}}%
 \let\@evenfoot\@oddfoot}

\@ifundefined{definecolor}{\usepackage{color}}{}
\definecolor{Wine}{rgb}{0.612, 0.192, 0.388}

\usepackage{marginnote}
\usepackage{lineno}

\@ifundefined{showcaptionsetup}{}{%
 \PassOptionsToPackage{caption=false}{subfig}}
\usepackage{subfig}
\makeatother

\newcommand\inputpgf[2]{{
\let\pgfimageWithoutPath\pgfimage
\renewcommand{\pgfimage}[2][]{\pgfimageWithoutPath[##1]{#1/##2}}
\input{#1/#2}
}}

\newlength\figureheight
\newlength\figurewidth

\title{Gap junction plasticity as a mechanism to regulate network-wide oscillations}

\author[1]{Guillaume Pernelle}
\author[1]{Wilten Nicola}
\author[1]{Claudia Clopath}
\affil[1]{Department of Bioengineering, Imperial College London}

\pgfplotsset{compat=1.14}
\begin{document}

\maketitle

\begin{abstract}

Cortical oscillations are thought to be involved in many cognitive functions and processes.  Several mechanisms have been proposed to regulate oscillations.
One prominent but understudied mechanism is gap-junctional coupling. 
Gap junctions are ubiquitous in cortex between GABAergic interneurons. 
Moreover, recent experiments indicate their strength can be modified in an activity-dependent manner, similar to chemical synapses.  We hypothesized that activity-dependent gap junction plasticity acts as a mechanism to regulate oscillations in the cortex.   We developed a computational model of gap junction plasticity in a recurrent cortical network.  We showed that gap junction plasticity can serve as a homeostatic mechanism for oscillations by maintaining a tight balance between two network states: asynchronous irregular activity and synchronized oscillations. This homeostatic mechanism allows for robust communication between neuronal assemblies through two different mechanisms:  transient oscillations and frequency modulation. This implies a direct functional role for gap junction plasticity in information transmission in cortex.  

\end{abstract}

\section{Introduction}
Oscillatory patterns of neuronal activity are reported in many brains regions with frequencies ranging from less than one Hertz to hundreds of Hertz.
These oscillations are often associated with cognitive phenomena such as sleep or attention. Local field potential measurements in the neocortex and thalamus show the prevalence of delta oscillations (0.5-4Hz) and spindle oscillations (7-15Hz) during sleep (\cite{Timofeev2012}). Theta oscillations (4-10Hz) are also reported in hippocampus and other brain regions (\cite{Buzsaki2002}). Gamma oscillations (30-100Hz) observed in the cortex are thought to be involved in attention
(\cite{Fries2001,Gregoriou2009,Vinck2013,Rouhinen2013}), perception
 (\cite{Rodriguez1999,Melloni2007}) and coordinated motor
output (\cite{Baker2007,Omlor2007}). Thus, at the minimum, oscillations are present during the normal functioning of neural circuits.

However, oscillations are also associated with pathological circuit dynamics, such as hyper-synchronous activity during epileptic seizures (\cite{Fisher2005}). 
Altered gamma-frequency synchronizations may also be involved in cognitive abnormalities such as autism (\cite{Orekhova2007}) or schizophrenia (\cite{Lewis2005}). Thus, given both the functional and pathological effects of oscillations, a homeostatic mechanism is necessary to regulate oscillatory behavior.

Several mechanisms can lead to the emergence of oscillations. They can arise in homogeneous population of excitatory neurons, where the positive feedback loop of excitation is only limited by the refractoriness of the neurons (\cite{Brunel2000c}). 
Alternatively, oscillations can also arise in a coupled network of excitatory and inhibitory neurons, where the excitatory and inhibitory neurons burst in opposing phase. (\cite{Sanchez-Vives2000,Haider2006,McCormick2015,Buzsaki2015,Veit2017}). 
Finally, gap junctions between inhibitory neurons promote synchronous oscillatory patterns (\cite{Traub2001,Pfeuty2003,Kopell2004,Connors2004a,Tchumatchenko2014}).

The inhibitory network oscillations primarily involve fast-spiking interneurons. These neurons represent a large proportion of GABAergic interneurons (\cite{Kawaguchi1997}). They are the main cells targeted by thalamocortical synapses transmitting sensory information to the cortex (\cite{Gibson1999}). 
They are coupled via chemical synapses and gap junctions. Gap junctions are mostly found between neurons of the same class and rarely otherwise (\cite{Galarreta1999,Gibson1999,Chu2003}). 
Moreover, there is evidence of the critical role of fast-spiking parvalbulmin (FS) interneurons in the emergence of cortical gamma activity in the cortex of rodents in response to sensory stimuli (\cite{Whittington2011,Bartos2007b,Cardin2009,Sohal2009}). 

Two main properties of FS interneurons have been found critical in the existence of gamma oscillations.
Firstly, FS interneurons selectively amplify gamma frequencies (\cite{Cardin2009}). 
Secondly, gap junctions between inhibitory interneurons (\cite{Galarreta1999}) have been shown to enhance synchrony (\cite{Gibson1999,Tamas2000,Hormuzdi2001,Buhl2003,Traub2004,Ostojic2009,Wang2014,Tchumatchenko2014,Robinson2017}). 
A computational model with both properties, inhibitory neurons with subthreshold resonance, connected by gap junctions, has been shown to support gamma oscillations (\cite{Hutcheon2000,Pike2000,Fellous2001,Tateno2004,Manor1997,Tchumatchenko2014}).

Recently, gap junction plasticity has been experimentally demonstrated \cite{Cachope2007a,Wang2015a,Turecek2014,Turecek2016}. For example, the gap junctions between rod cells in the retina can vary their conductance during day and night cycles (\cite{Jin2016}).
Moreover, they can experience bidirectional long-term plasticity in an activity-dependent manner (\cite{Cachope2012,Haas2016}). High frequency stimulation of a coupled pair thalamic reticular nucleus neurons leading to burst firing induces gap junction long-term depression (gLTD) (\cite{Haas2011}). \cite{Sevetson2017} show that the pathways leading to gLTD are calcium-dependent which suggest that gap junction long-term potentiation (gLTP) could also be the result of spiking activity.


Given the existence of gap junction plasticity and the necessity of a homeostatic mechanism for regulating oscillations, we wondered whether gap junction plasticity can regulate network-wide gamma oscillations in cortex.
To that end, we developed a computational model of a network of excitatory and FS inhibitory neurons. 
As demonstrated analytically by~\cite{Tchumatchenko2014}, we observed two
different network behaviors depending on the gap junction strength. 
For weak gap junction strength, the network exhibits an asynchronous regime, whereas for strong gap junctions, the network undergoes gamma oscillations with bursting activity. 
We then modelled the gap junction plasticity observed by~\cite{Haas2011} showing that bursting activity lead to gLTD. The plastic network sets itself at the transition between the asynchronous regime, where sparse spiking dominates, and the synchronous regime, where network oscillations dominate and burst firing prevails. 
Thus, our model shows that gap junction plasticity maintains the balance between the asynchronous and synchronous network states. 
We then show that the network allows for transient oscillations driven by external drive. This demonstrates that transient, plasticity regulated oscillations can efficiently transfer information to downstream networks. Finally we show that gap junction plasticity mediates cross-network synchronization and allows for robust information transfer trough frequency modulation. Critically, gap junction plasticity allows for the recovery of oscillation mediated information transfer in the event of partial gap junction loss. 

\section{Results}

\subsection*{Network synchrony depends on gap junctions strength.}

To study the effect of gap junction plasticity, we developed a network of coupled inhibitory and excitatory neurons in the fluctuation-driven state (Figure 1A). 
The Izhikevich model was used for the inhibitory neuron population to fit the fast-spiking inhibitory neuron firing pattern (\cite{Izhikevich2007}). Excitatory neurons are modelled by leaky integrate-and-fire models.
As in~\cite{Tchumatchenko2014}, the excitatory neurons act as low pass-filters for their inputs while the FS neurons have a sub-threshold resonance in the gamma range (\cite{Hutcheon2000,Pike2000,Fellous2001,Tateno2004,Manor1997}.) 
To demonstrate this, we injected an oscillatory current of small amplitude in a single cell and recorded the amplitude response for different oscillatory frequencies. 
Excitatory neurons better respond to low frequency inputs, while FS neurons respond maximally for gamma inputs (Figure 1B). This is in line with the experimental evidence of Cardin et al. showing that FS-specific light stimulation amplifies gamma-frequencies (\cite{Cardin2009}). 

All neurons have chemical synapses but only inhibitory neurons are also coupled via gap junctions (Figure 1A). 
The gap junctions are modelled such that a voltage hyperpolarization (depolarization) in one neuron induces a voltage hyperpolarization (depolarization) in the connected neuron. 
The ratio between the two voltage deviations corresponds to the gap junction strength $\gamma$ (Figure 1C). 
Moreover, when one neuron spikes, it emits a spikelet in the coupled neuron. 
We model this by a positive inhibitory to inhibitory electrical coupling, which we add on top of the negative inhibitory to inhibitory chemical coupling (see \textit{Materials and Methods}).

In order to understand the effects of gap junction plasticity, we first considered the network without plasticity.
We first explored the network behavior for different values of the mean gap junction strengths $\gamma$ and mean external drive to the inhibitory neurons $\nu_{I}$. 
As demonstrated by~\cite{Tchumatchenko2014}, our network exhibits two regimes (Figure 1D):
an asynchronous irregular (AI) regime and a synchronous regular regime (SR).  
The AI regime occurs for networks with weak external drive and weak gap junctions. 
In this regime the network is in the fluctuation driven regime so that the neurons spike due to variations in their input. 
The SR regime occurs for strong external drive and strong gap junctions. This regime leads to the emergence of gamma oscillations. 
Mathematically, the network undergoes a Hopf bifurcation (\cite{Ostojic2009,Tchumatchenko2014}). 
The oscillations arise as the network directly inherits the resonance properties of the individual neurons. 
This is mediated through the gap junction coupling which effectively allows positive coupling through their spikelets. 
Moreover, the gap junctions reduce subthreshold voltage differences between neurons which promotes synchrony. 
The excitatory neurons are not necessary for the oscillations but they amplify the dynamics (see~\cite{Tchumatchenko2014} for mathematical derivations). 
When placed in the SR regime, the network oscillates in the gamma-range at a frequency near the single neuron resonance frequency (Figures 1E-F). 
In addition, we observe that the spiking activity is characteristic to the network regime, with bursting activity in the synchronous regime and spikes in the asynchronous regime (Figures 1G-I).

To summarize, increased gap junction coupling and input drive into the network promotes gamma oscillations. 
To explain the relationship between network activity and gap junction plasticity, we first model the simplest case of plasticity between a pair of electrically coupled neurons. 
We then extend the plasticity rule to a population of neurons and investigate the effects on the network dynamics.


\subsection*{Model of gap junction plasticity: bursting induces gLTD, spiking gLTP.}

To determine how gap junction plasticity can alter network dynamics, we developed a model of the plasticity based on experimental observations. \cite{Haas2011} have shown that bursts in one or both neurons in an electrically coupled pair lead to long-term depression (gLTD). 
Therefore, we modeled gLTD as a decrease in the gap junction strength that is proportional to the amount of bursting. The constant of proportionality, $\alpha_{gLTD}$ serves as the learning rate. To infer $\alpha_{gLTD}$, we reproduced the bursting protocol in Haas et al., where a neuron bursting for a few milliseconds, 600 times for 5 minutes, leads to 13\% decrease (Figure 2A).

Activity-dependent gap junction long-term potentiation (gLTP) has not been reported experimentally yet in the mammalian brain. 
There is evidence for activity dependent short-term potentiation in vertebrates~\cite{Pereda1996,Cachope2012}. However, without potentiation, all gap junctions would likely become zero with time. 
To address this concern, we assume that gap junctions can undergo gLTP and we modeled it such that single spikes induce gLTP by a constant amount given by the potentiation learning rate $\alpha_{gLTP}$ (Fig 2B, first half).

\subsection*{Gap junction plasticity regulates network-wide oscillations}
Our plasticity model therefore potentiates gap junctions under spiking activity and depresses under bursting activity. 
Therefore, we wondered how gap junction plasticity can alter network dynamics. We previously quantified the amount of spiking versus bursting in our network for different levels of fixed gap junction strength and mean drive. For low levels of both, the network is spiking whereas for high levels of both the network is bursting.
The spiking to bursting transition (Figure 1G) corresponds to the bifurcation (Figure 1D) from asynchronous irregular to synchronous oscillations at gamma frequency.  
When inhibitory neurons are oscillating, they fire a burst of spikes at the peak of the oscillations (Figure 1I, $\gamma=5$).
Therefore, when gap junctions are plastic, the network steady state can be found on the side of the bifurcation that balances the amount of potentiation due to spiking activity with the amount of depression due to bursting activity. 
The depression learning rate is inferred from Haas et al., while the potentiation learning rate is left as a free parameter. 

We found that a strong relationship exists between gap junction plasticity and network synchrony. 
When the network is in the AI regime, characterized by low prevalence of bursting activity, gap junction potentiation dominates. 
However, for a strong mean coupling strength, the emergence of oscillations is associated by high bursting activity which leads to depression of the gap junctions. 
Therefore gap junction plasticity in our network maintains a tight balance between asynchronous and synchronous activity. Depending on the value that we choose, the position of the plasticity fixed point lies either in the asynchronous regime, for lower $\alpha_{gLTP}$ values, or in the synchronous regime for higher $\alpha_{gLTP}$ values. 

\subsection*{Gap junction plasticity allows for sparse but salient information transfer}
We wondered how gap junction plasticity would interact with time-varying inputs. For the following experiment we consider the gap junction plasticity fixed point to be in the asynchronous irregular regime. 
First, we let the network reach its steady state with a low level of drive (Figure 2E, beginning). 
As previously observed, the mean gap junction strength reaches a value which sets the network near the AI/SR transition. 
Then, we proceeded by injecting an additional constant current to the network. 
This new current base-line induced network level oscillations (Figure 2E, transition). 
However, over time the mean gap junction strength decays due to the gap junction plasticity mechanism.
This gap junction depression is followed by a loss of synchrony and the network reaches its new steady state (Figure 2E, end), again near the border of asynchronous and synchronous regimes.

We measured the response of read-out neurons which receive projections from the excitatory and inhibitory neurons in our network. 
At the onset of the current step, the network underwent transient oscillations. When the gap junctions are plastic, the downstream neurons increase their spiking activity only for a few hundred milliseconds during the transient oscillations and then became almost quiescent again (Figure 2F, second panel). This contrasts with the simulation of a static network where the downstream keep a high firing rate (Figure 2F, third panel).

These results suggest that synchronous activity is a powerful signal to provoke spiking in downstream neurons. But oscillations and high firing rates of downstream neurons are also metabolically costly (\cite{Kann2011}). With transient oscillations however, the downstream neurons only sparsely fire when the stimulus changes but not when it is predictable. Thus, the regulation of oscillations mediated by gap junction plasticity allows for sparse but salient information transfer.

\subsection*{Gap junction plasticity enhances the ability of sub-populations of neurons to synchronize.}

We now sought to study what could be the functional implications of a plasticity fixed point in the SR regime.
Synchronization between networks is considered as one of the possible mechanism of information transfer (\cite{Salinas2001,Tiesinga2001,Fries2005a,Fries2015}). 
We wondered whether gap junction coupling could mediate cross-network synchronization, and how gap junction plasticity would regulate this synchronization. 
To test this hypothesis, we considered two subnetworks having different oscillation frequencies and coupled by gap junctions (Figure 3A).
A fast network oscillates at a gamma frequency and therefore is called the gamma-network. 
Then, a slow-network oscillates at a slower frequency as the membrane time constant of its inhibitory neurons is chosen to have a larger value.
Indeed, previous analyses show that the network frequency in our model is inherited from the single neuron resonance frequency of inhibitory neurons (\cite{Tchumatchenko2014,Chen2016}). 
As a result, increasing the membrane time constant of the inhibitory neurons results in a decrease of the network oscillation frequency (Figure 3B-D).
Cross-network gap junctions reduces the frequency difference between the gamma- and slow-network (Figures 3E and 3I) and larger differences of subnetwork resonant frequencies require a larger number of cross-network gap junctions for the networks to oscillate in harmony. 
Their common frequency lies between the resonant frequencies of each network, would they be decoupled. 
Importantly, cross-network synchronization requires the subnetworks to be in phase. If the gamma- and slow-network do not share enough gap junctions, there is no correlation in their population activities (Figure 3H), despite sharing chemical synapses and having a common oscillation frequency in some cases ([$\Delta f_{res}$=0;$s.GJs$=0] or [$\Delta f_{res}$=18;$s.GJs$=48] on Figure 3I).
However, for small differences in the subnetworks resonant frequency $\Delta f_{res}$, increasing the number of shared gap junctions induces the oscillations to lock together. 
The networks oscillate in phase (Figure 3H, end of first row) as reflected in their correlation (Figure 3I, dark red area).
In summary, two networks in the SR regime with different resonance frequencies and/or out-of-phase can synchronize if they are coupled by gap junctions. Furthermore, a large number of shared gap junctions is required for large differences of resonant frequency.  

As gap junctions can synchronize two oscillating populations of neurons, we wondered whether the same synchronization would occur with one population in the AI regime. First, we initialized the gamma-network in the AI regime while the SN was initialized in the SR regime (Figure 4A). After coupling the gamma- and slow-network together, we found that, while the oscillation frequency of the gamma- and slow-network matched (Figure 4B), the two networks could not synchronize. The networks were always out-of-phase with very weak correlation between the population activities (Figure 4C, 4D). The results were similar if the gamma- and the slow-network were initialized in the reverse synchronous and asynchronous parameter regimes, respectively (not shown). Cross-network synchronization is not robust when one network is not oscillatory.   

Given these constraints on cross-network synchronization, we wondered if gap junction plasticity could remedy the situation and allow for robust cross-network synchronization. 
To test this hypothesis, we repeated the simulation protocols with the gamma- and slow-network initialized in the asynchronous and synchronous regimes (respectively) and with plastic gap junctions. 
Here we considered the case where the gap junction plasticity steady state lies in the synchronous regime.
As shown previously, gap junction plasticity regulates oscillations such that the network in the asynchronous irregular regime transitions to the oscillatory regime (Figure 4E). The oscillation frequencies of these two networks match (Figure 4G).  
Strikingly, even with a large resonant frequency difference, the gamma- and slow-network now synchronize with through a small number of shared gap junctions (4G, 4H). This indicates that gap junction plasticity allows for cross-network synchronization that is robust to the underlying neuronal parameters for small numbers of shared gap junctions.


\subsection*{Gap junction plasticity allows for robust information transfer.}

We hypothesized that cross-network synchronization mediated by plasticity allows information transfer. To investigate this, we considered a similar network architecture as previously studied, with two networks, an input-network and an output-network. The input-network receives an input projected by random weights to its neurons. The output-network is connected to the input-network with a small number of gap junctions and inhibitory chemical synapses. 

First, to demonstrate the information transfer capability of the network, we consider static gap junctions with oscillatory inputs to the IN.
The stimulus information is transmitted to the output-network via the frequency modulation of the synchronized oscillations and not by spike transmission nor amplitude modulation (Figures 5A-D).
When sharing gap junctions, the input- and output-network synchronize together (Figure 5A) and their spiking activity is locked (Figure 5B). As the amplitude of the input signal increases, the spiking activity increases in the input-network but not in the output-network (Figure 5C). 
For a network in the SR, there is a positive correlation between the signal amplitude and the network oscillation frequency (Figures 1E and 5D). 
This frequency modulation is transferred from the input- to the output-network. 
Thus, the input amplitude can be estimated from the oscillation frequency of the ON, despite the absence of chemical synapses between the input-network and the output-network (Figure 5E).  
However, this synchrony code is only possible for signals below a certain frequency (Figures 5F-G). Indeed, the instantaneous oscillation frequency is estimated by measuring the period between consecutive peaks of the population activity. For example, oscillations at 50 Hz have a period of 20 ms. Variations happening within those 20 ms are compressed to a single period value and thus are not transferred via frequency modulation. Mechanisms for estimating the input value from the oscillation frequency of the output-network are discussed further in the methods section.
Finally, we tested if this synchrony code was valid for non-oscillatory signals. We found that non-oscillatory, slowly varying random signals could also be robustly transmitted from the input- to the output-network with gap junction coupling (Figure 5I).

As gap junction plasticity can regulate oscillations, we tested whether the plasticity can make this synchrony code robust to parameter variations or potential gap junction loss. 
First, as previously shown, gap junction plasticity enhances the ability of networks to synchronize. 
If initialized in the AI regime and with static gap junctions, there is no information transfer via frequency modulation (Figure 5J, left panel). However, with plasticity the oscillations are regulated and the network synchrony is recovered which results in successful information transfer (Figure 5J left panel).
A critical amount of oscillation power and a critical number of shared gap junctions are required for information transfer, after which increasing each of them does not yield significant improvement (Figure 5J). Furthermore, we studied whether gap junction plasticity could restore information transfer if gap-junctions were deleted.  
While there is loss in the quality of the transfer when static gap junction are removed, plastic gap junctions maintain the quality of the transfer by increasing the strength of the remaining gap junctions. 
This mechanism compensates for the missing gap junctions (Figures 5 J-K). 

To summarize, gap junction plasticity expands the necessary conditions for information transfer. It regulates oscillations, and by promoting phase-locking of oscillations, it contributes to the propagation of information to downstream networks. Finally, if some gap junctions are failing, due to protein turnover perhaps, the remaining ones can increase their strength through plasticity.  This helps to maintain accurate information transfer.

\section{Discussion}

Our modelling study tested whether gap junction plasticity can regulate gamma oscillations in cortical network models. Our findings suggest that gap junction plasticity can maintain a balance between synchronous regular and asynchronous irregular regimes. For strong electrical coupling, the network is in the oscillatory regime. The oscillations consist of synchronized bursting mediated by the inhibitory neuron network. These bursts trigger depression of the gap junctions (\cite{Haas2011}) allowing the network to leave the oscillatory regime and spike asynchronously. However, the irregular asynchronous regime is dominated by sparse firing, which we assume provides potentiation. Thus, the asynchronous irregular regime tends to potentiate gap junctions.  
Therefore, equilibria can be found on either side of the bifurcation, either in the asynchronous irregular or in the synchronous regular regime, depending on the plasticity learning rates. We demonstrate the functional role of plasticity in both cases. First, with equilibria in the AI regime, the network can respond to changes in input drives through transient oscillations. Those transient oscillations could serve as an energetically efficient way to transfer information to a downstream neuron. Second, with equilibria in the SR regime, the network oscillations can serve as the substrate for information routing between networks. These results demonstrate how gap junction plasticity can regulate oscillations to mediate information transfer between cortical populations of neurons.

\textbf{Gap junction coupling between interneurons affects network synchrony.}
Despite being less common than chemical synapses,  gap junctions are ubiquitous in the central nervous system. Example includes the inferior olivary nucleus (\cite{Sotelo1974,Llinas1974,Benardo1986}), the thalamic reticular nucleus (\cite{Landisman2002,Long2004}), the hippocampus (\cite{Jefferys1995,Hormuzdi2001}), the retina (\cite{Vaney2002,Jin2016}), the olfactory bulb (\cite{Zhang2003}), the locus coeruleus (\cite{Christie1989}), or also the neocortex (\cite{Sloper1972, Sloper1978}). 
Moreover, they drastically alter the firing activity of their connecting neurons (\cite{Haas2015,VanWelie2016}), as well as the network dynamics (\cite{Traub2001,Pfeuty2003,Kopell2004,Connors2004a,Tchumatchenko2014}). 
Furthermore, gap junctions between inhibitory interneurons are reported in many cortical regions where global oscillations of neural activity are observed (\cite{Galarreta1999,Deans2001,Pfeuty2003,Gibson2005}). 
These inhibitory neurons exhibit subthreshold resonance that amplifies a specific frequency range (\cite{Cardin2009}). 
Therefore, gap junction induced synchrony and inhibitory neurons frequency preference are a possible substrate for global oscillations in these cortical regions.
Our work is consistent with results of~\cite{Tchumatchenko2014,Chen2016} showing that together gap junction strength and subthreshold resonance of inhibitory neuron promote oscillations of neuronal activity.  

\textbf{Model of gap junction plasticity: bursts induce gLTD, spikes induce gLTP}
Recently,~\cite{Haas2011} reported the first experimental evidence of activity-dependent gLTD of gap junctions of interneurons in the thalamic reticular nucleus, even though the mechanism remains to be investigated (\cite{Szoboszlay2015}). Also \cite{Sevetson2017} found that calcium-regulated mechanisms support gap junction gLTD in the thalamic reticular nucleus. The mechanisms are similar to those observed for chemical synapse plasticity.
We designed a rule for activity-dependent gLTD consistent with those results. We assumed that a cortical fast-spiking interneuron would exhibit the same plasticity properties as a thalamic reticular neuron because gap junctions are mostly made from the connexin Cx36 throughout the central nervous system (\cite{Landisman2002,Rouach2002}). 
To our knowledge, there is no study yet on activity-dependent gLTP of gap junctions.
However recent studies (\cite{Wang2015a,Sevetson2017}) suggests that gLTD and gLTP share a common pathway. 
Therefore, we propose a rule for activity dependent gLTP, assuming that low frequency spiking activity leads to gap junction potentiation.

\textbf{Gap junction plasticity regulates oscillations and propagates transient information.}
Our model demonstrates that the regulation of oscillations is mediated by gap junction plasticity. Sparse firing in the AI regime leads to potentiation which increases the network synchrony, while bursting activity associated with the SR regime leads to depression. Our first hypothesis assumed that the plasticity fixed point is in AI regime. Thus, at the steady-state, gamma power is weak or non-existent. Evidence from (\cite{Tallon-Baudry1999a,Ray2015}) is consistent with our results. When no stimulus is provided or task required, electroencephalogram recordings show that power in the gamma-band is weak.
After the onset of a sensory stimulus, gamma oscillations can be detected in cortical areas. This has been reported for example with visual stimuli triggering gamma oscillations in the mouse visual cortex (\cite{Saleem2017}). 
In our model, the neurons oscillate transiently when receiving a constant external stimulation. This mechanism operates by crossing the bifurcation boundary between the AI and SR regime. However, over time the mean gap junction strength decays due to the additional bursting activity. The gap junction depression leads to a loss of synchrony and the network reaches its new steady state in the asynchronous regime again. Therefore we predict a loss in gamma power for sustained stimulus. A similar mechanism may be involved in the reduction of gamma oscillation induced by slow smooth movements (\cite{Kruse1996,Tallon-Baudry1999}).

We wondered what could be the functional role of this transient oscillatory regime. Projecting the excitatory activity of our network model to downstream neurons revealed that they fire sparsely, for a short duration after stimulus onset, and are quiescent otherwise. Thus, gap junction plasticity could efficiently encode the change in incoming stimuli. This could allow for energy conservation as oscillations are energetically expensive (\cite{Kann2011}). Moreover, \cite{Palmigiano2017} show that cortical circuits near the onset of oscillations could promote flexible information routing by transient synchrony.

\textbf{Plastic gap junction coupling for robust information routing}.
The role of gamma oscillations is highly debated (\cite{Ray2015}). 
They could play no role and simply be a marker of the excitation-inhibition interaction. However others studies suggest they could be involved in information transfer. It is thought that retinal oscillations carry information to the visual cortex (\cite{Koepsell2009}). 
Moreover they could serve as inter-area communication by promoting coherence in neural assemblies which would align their windows of excitation. 
This would allow for effective spike transmission (\cite{Ray2015,Fries2005a,Bosman2012a}). 
Furthermore, \cite{Roberts2013a} observed high gamma coherence between layers 1 and 2 of macaque's visual cortex by dynamic frequency matching. 
Here, we demonstrate one potential mechanism for information transmission through gamma oscillations. 
Our networks make use of gamma frequency modulation to transmit information in a robust manner, similar to the principle used for FM radio broadcasting. 
The amplitude of the input signal modulates the oscillation frequency, which increases almost linearly with the amplitude. 
Our model demonstrates that gap junction plasticity robustly mediates network oscillations and cross-network synchronization. If some gap junctions are removed, the remaining gap junctions become stronger and compensate for the missing ones. Thus, gap junction plasticity insures the phase-locking of the coupled network and it allows for information routing. In particular, there is evidence suggesting that gap junctions could promote long-distance signaling by implementing frequency modulation of calcium waves in astrocytes (\cite{Goldberg2010}). 
Moreover, correlation was found during gamma activity between amplitude and frequency modulation of local field potential of CA3 pyramidal neurons of anesthetized rats (\cite{Atallah2009}). 

Failure to regulate oscillations, could be the origin of several cognitive pathologies. 
Disruption of brain synchrony in the inferior olive is thought to contribute to autism due to the loss of coherence in brain rhythms (\cite{Welsh2005}). Excess of high frequency network wide oscillations in the cortex have been observed to also correlate with autism in young boys (\cite{Orekhova2007}). The inferior olive differs for its density of gap junction being the highest in the adult brain (\cite{Llinas1974,Sotelo1974}). It may be involved in the generation of tremors in Parkinson's disease, however the severity of induced tremors in Cx36 knockout mice remained the same as in wild-type mice (\cite{Nakase2004,Long2002}). This could be due to gap junctions made from other connexins (such as Cx43) taking over the knocked-out ones.

Recent studies highlight the critical role of gap junction plasticity in efficient cognitive processing. As experimental and computational techniques improve, new efforts can further unveil their properties and expand our understanding of cortical functions. Our computational model shows that gap junction activity-dependent plasticity may play an important role in network-wide synchrony regulation.

\section{Methods}

We consider a network with $N_I$ inhibitory neurons (20\%) and $N_E$ excitatory neurons (80\%) with all-to-all connectivity (Figure 1A). 
Inhibitory neurons are modelled by an Izhikevich model and excitatory neurons by a leaky integrated-and-fire model (LIF) (\cite{Izhikevich2003,Izhikevich2007}). 
The simulation time-step is $dt=0.1$ ms. Inhibitory neurons are connected by both electrical and chemical synapses, whereas excitatory neurons have only chemical synapses. 
We designed a novel plasticity model for activity dependent plasticity of gap junctions and we investigated its impact on network dynamics and function. 
We then investigated the dynamics of two networks coupled by chemical and electrical synapses. 
We use a decoder to quantify the effects of gap junction plasticity on information transfer. 
The model is written in Python and takes advantage of the tensorflow library that leverages GPU parallel processing capabilities (\cite{Abadi2016}).

\subsection*{Neuron model} \label{s:izh}
We model Fast Spiking (FS) interneurons with Izhikevich type neuron models (\cite{Izhikevich2007}). 
This model offers the advantage to reproduce different firing patterns as well as a low computational cost (\cite{Izhikevich2004}).  
The voltage $v$ follows 
\begin{equation} \tau_v \dot{v} =  (v-v_{ra})(v-v_{rb}) - k_u u + R I,
\end{equation}
\begin{equation} \tau_u \dot{u} = a(v - v_{rc}) - u, \end{equation}
combined with the spiking conditions,
	\begin{equation}
		\mbox{
			if $v \geq v_{threshFS}$,
			then}
			\left\lbrace{ \begin{array}{l} v \leftarrow v_{resetFS} \\ u \leftarrow  u+b.
\end{array}} \right.  \end{equation}
where $\tau_v$ is the membrane time constant, $v_{ra}$ is the membrane resting potential, $v_{rb}$ is the membrane threshold potential, $k_u$ is the coupling parameter to the adaptation variable $u$, $R$ is the resistance and $I$ is the current. 
The adaptation variable $u$ represents a membrane recovery variable, accounting or the activation of K$^+$ ionic currents and inactivation of Na$^+$ ionic currents. It increases by a discrete amount $b$ every time the neuron is spiking and its membrane potential crosses the threshold $v_{threshFS}$.
It provides a negative feedback to voltage $v$.
$\tau_u$ is the recovery time constant, $a$ is a coupling parameter, $v_{resetFS}$, $b$ and $v_{rc}$ are voltage constants.

For the FS neurons, we chose the membrane potential reset $v_{resetFS}$ and the spike-triggered adaptation variable $b$ to account for the onset bursting activity observed \textit{in vivo}. 
Modifying $k_u$, $v_{ra}$, $v_{rb}$ and $v_{rc}$ was sufficient to observe the emergence of a resonance frequency.
We set the time constant $\tau_u$ to obtain a resonance frequency of 45 Hz, which is in the same range as observed \textit{in vivo} by~\cite{Cardin2009} (Figure 1B). 
To measure the subthreshold resonant property (Figures 1B, 3B and 3D), we recorded the amplitude of the neuronal membrane potential $V_E$ in response to different oscillation frequencies $f$ of
low level sinusoidal currents $I(t)= I_0 cos(2 \pi f t) $ (with $I_0=0.01$ pA). 
We then normalized the amplitude response as follow
\begin{equation}
R_E(f) = \frac{|V_E(I_0cos(2\pi f t))|}{max(|V_E(I_0cos(2\pi f t))|)},
\end{equation}
for frequencies between 0 and 1kHz. The $|$ $|$ denotes the absolute value. 

To model regular spiking excitatory neurons, we chose a leaky integrate-and-fire model, 
\begin{equation} \tau_m  \dot{v} =  -v + R_m I,
\end{equation}
where $\tau_m$ is the membrane time constant, $v$ the membrane potential, $I$ the current and $R_m$ the resistance.
Spikes are characterized by a firing time $t_f$ which corresponds to the time when $v$ reaches the threshold $v_{threshRS}$.
Immediately after a spike, the potential is reset to the reset potential $v_{resetRS}$.

\subsection*{Network.}\label{sec:network}

In the single network model (Figures 1 and 2), each neuron is connected to all others by chemical synapses, but in addition, inhibitory neurons are connected via  electrical synapses to all other inhibitory neurons, as in~\cite{Tchumatchenko2014}. 
Thus, the current each individual neuron $i$ receives can be decomposed in four components 
\begin{equation} 
I _{i}(t)= I^{spike}_{i}(t) +I^{gap}_{i}(t) + I^{noise}_{i}(t) + I^{ext}_{i}(t), 
\end{equation}
where $I^{spike}_{i} = I^{chem}_{i} + I^{elec}_{i}$ is the current coming from the transmission of a spike via electrical (i.e. spikelet) and chemical synapses, $I^{gap}_{i}$ is the subthreshold current from electrical synapses (for inhibitory neurons only), $I^{noise}_{i}$ is the noisy background current and $I^{ext}_{i}$ characterizes the external current.  
The current due to spiking $I^{spike}_{i}$ on excitatory neurons is given by 
\begin{equation}
	I^{spike}_{i}(t) = W^{IE} 
	\sum_{	
	\substack{
				j=1\\
				j \neq i}
		}^{N_I}
		\sum_{
			t_{jk}<t
		} 
			\exp{\left(
				-\frac{t-t_{jk}}{\tau_{I_I}}
				\right)}
			\\
			+
	W^{EE}  \sum_{	
	\substack{
				j=1\\
				j \neq i}
		}^{N_E}
		\sum_{
			t_{jk}<t
			} 
				\exp{\left(-\frac{t-t_{jk}}{\tau_{I_E}}\right)}.
				\end{equation}
				
The current $I^{spike}_{i}$ into inhibitory neurons are
\begin{equation}
					I^{spike}_{i}(t) =
		\sum_{	
	\substack{
				j=1\\
				j \neq i}
		}^{N_I}
		\sum_{
			t_{jk}<t
		} W^{II}_{ij} 
			exp{\left(
				-\frac{t-t_{jk}}{\tau_{I_I}}
				\right)}
			\\
			+
	W^{EI} 	\sum_{	
	\substack{
				j=1\\
				j \neq i}
		}^{N_E}
		\sum_{ 
			t_{jk}<t
			} 
				\exp{\left(-\frac{t-t_{jk}}{\tau_{I_E}}\right)},
\end{equation}

where $W^{\alpha\beta}$ is the coupling strength from population $\alpha$ to population $\beta$ with $\{\alpha,\beta\} = \{E,I\}$.  
Finally, $W^{II}_{ij} = W^{II,c} + W^{II,e}_{ij}$ is the inhibitory to inhibitory coupling between neuron $i$ and $j$, consisting of the chemical synaptic strength $W^{II,c}$ and $W^{II,e}_{ij}$ the electrical coupling for supra-threshold current, also called the spikelet. 
We model the contribution of the spikelet as a linear function of the gap junction coupling $W^{II,e}_{ij} = k_{spikelet} * \gamma_{ij}$, where $\gamma_{ij}$ is the gap junction coupling between neurons $i$ and $j$.
Note that $W^{EE}$, $W^{EI}$, $W^{IE}$, $W^{II,c}$ are identical among neurons, but $W^{II}_{ij}$ varies as the spikelet contribution depends on the coupling strengths $\gamma_{ij}$, which can be plastic. 
We also modeled the network with chemical weights following a log-normal distribution, which yielded similar results (data not shown).

We represent post-synaptic potential response to a chemical or electrical spike with an exponential of the form 
$\exp{\left(-\frac{t-t_{jk}}{\tau_{I_{\alpha}}}\right)}$ for $t>t_{jk}$. The excitatory and inhibitory synaptic time constants are $\tau_{I_E}$ and $\tau_{I_I}$ respectively and $t_{jk}$ represents the $k^{th}$ firing time of neuron $j$.  	

In between spikes, for every pair of inhibitory neurons $i,j$, the gap junction
mediated subthreshold current $I^{gap}_{i}$ is characterized by

\begin{equation}
		I^{gap}_{i}(t) =
		\sum_{	
	\substack{
				j=1\\
				j \neq i}
		}^{N_I}
		I^{gap}_{ij}(t) =
		\sum_{	
	\substack{
				j=1\\
				j \neq i}
		}^{N_I}
		\gamma_{ij} (V_{j}(t) - V_{i}(t)),
\end{equation} 
where $\gamma_{ij}$ is the coupling coefficient between inhibitory neurons $i$ and $j$ of respective membrane
potential $V_{i}$ and $V_{j}$. In our model, we suppose that gap junctions are symmetric with $\gamma_{ij} = \gamma_{ji}$. 
Gap junctions are initialized following a log-normal distribution with the location parameter $\mu_{gap}=1+ln(\gamma / N_I)$ and the scale parameter $\sigma_{gap}=1$.

Neurons also receive the current $I_{noise}$ which is a colored Gaussian noise
with mean $\nu_{I}$, standard deviation $\sigma_{I}$ and $\tau_{noise}$ the time
constant of the low-pass filtering

\begin{equation} 
	\tau_{noise}\dot{s}(t)=-s(t)+\xi(t) 
\end{equation} 
and
\begin{equation} 
	I^{noise}(t) = \sqrt{2\tau_{noise}}s(t)\sigma_{I} + \nu_{I},
\end{equation} 
with $\xi$ is drawn from a Gaussian distribution with unit standard deviation and zero mean.

\subsection*{Plasticity model of gap junctions.} 
Our plasticity model is decomposed into a depression ${\gamma}^{-}$ and a potentiation 
term ${\gamma}^{+}$. 

\subsection*{gLTD: depression of the electrical synapses for high frequency
activity}  \label{sec:LTD}~\cite{Haas2011} showed that bursting activity of both neurons or one of the two neurons leads to long-term depression (gLTD) of the electrical synapses. 
To capture this effect in our model, we first defined a variable $b_{i}$ which is a low-pass filter of the spikes of neuron $i$
\begin{equation}\label{eq:bursting} 
	\tau_{b} \dot{b_{i}}(t) = -b_{i}(t) + \tau_{b} \sum_{t_{ik}<t}  \delta(t-t_{ik}), 
\end{equation}

where $\delta$ is the Dirac function and $\tau_{b}=8$ ms is the time constant. When $b_{i}$ reaches a value of
$\theta_{burst}=1.3$, this indicates that two or more spikes happened within a short time interval. Therefore, burstiness of neuron $i$ is characterized by $H(b_{i}-\theta_{burst})$ where $H$ is the Heaviside function that returns 1 for positive arguments and 0 otherwise.

In our simplified model, we consider that the individual electrical coupling coefficient $\gamma$ between neurons are non-directional. Every time the interneurons burst, the gap junctions undergo depression,

\begin{equation}
\label{eq:ltd} \dot{\gamma}^{-}_{ij}(t)  = \dot{\gamma}^{-}_{ji}(t)  = -\alpha_{LTD}  [H(b_i(t)-\theta_{burst}) + H(b_j(t)-\theta_{burst})] , 
\end{equation} 

where $\alpha_{LTD}$ is the learning rate.

We fitted $\alpha_{LTD}$ to the data by implementing the stimulation protocol used in~\cite{Haas2011}.
We applied a constant current injection of 300 pA for 50 ms every 0.5 s (2 Hz) and of -80 pA the rest of the time, to maintain the membrane potential at -70 mV. 
This protocol lasts for 5 minutes. We estimate $\alpha_{LTD} = 1.569*10^{-5}$ $\Omega^{-1}s^{-1}$ by such that it leads to a depression of 13\% of the gap junction strength at the end of the stimulation protocol, as reported by Haas et al.

\subsection*{gLTP: potentiation of the electrical synapses for low frequency activity.} \label{sec:LTP} 
If gap junctions were only depressed, they would decay to zero after some time. Therefore, there is a need for gap junction potentiation. However, no activity dependent mechanisms was reported yet in the experimental literature, but \cite{Cachope2012,Wang2015a,Sevetson2017} suggest that the calcium-regulated mechanisms leading to long-term depression could be involved in potentiation as well. Therefore, in our model, we assume that spiking leads to long-term potentiation of the gap junction (gLTP) in contrast to bursting leading to gLTD. 

We consider two gLTP rules. The first has a soft bound, i.e. the magnitude of modification is proportional to the difference between the gap junction value and a baseline coupling strength $\gamma_b$ 

\begin{equation} \label{eq:ltp-soft} 
	\dot{\gamma}^{+}_{ij}(t) =\dot{\gamma}^{+}_{ji}(t)  =  \alpha_{LTP}  \left(\frac{\gamma_b - \gamma_{ij}(t)}{\gamma_b}\right) [\mathrm{sp}_i(t) + \mathrm{sp}_j(t)]. 
\end{equation}

where $\alpha_{LTP}$ is the learning rate and $\mathrm{sp}_i(t) = \sum_{t_{ik} < t} \delta(t-t_{ik})$ equals to 1 if neuron $i$ is spiking, and 0 otherwise. 
This softbound approach let us choose $\alpha_{LTP}$ and $\gamma_b$ such that the steady-state of the plasticity is found in asynchronous regime.

We also consider a gLTP rule without softbound, as following
\begin{equation} \label{eq:ltp-nosoft} 
	\dot{\gamma}^{+}_{ij}(t) =\dot{\gamma}^{+}_{ji}(t)  =  \alpha_{LTP} [\mathrm{sp}_i(t) + \mathrm{sp}_j(t)]. 
\end{equation}

While both rules can lead to plasticity fixed points in the synchronous regime, the first rule has the advantage to be more robust for obtaining fixed points in the asynchronous regime, as the potentiation decrease towards 0 approaching to a baseline coupling strength. Therefore we choose the softbound rule while considering plasticity fixed point in the asynchronous regime and we chose the second rule otherwise.

\subsection*{Quantification of network spiking activity.}

To estimate the plasticity direction for different value of external input $\nu$ and gap junction strength $\gamma$, we observe the activity of the network (without plasticity) in a steady state over a duration $T = 6$ s.
For a chosen tuple $(\nu;\gamma)$, we average over time and over neurons the bursting and spiking activity
\begin{equation}
\label{eq:avg1} 
	A_{bursting} = \frac{1}{T}\int_0^T \frac{1}{N_{I}} \sum_{i=1}^{N_I} [H(b_i(t)-\theta_{burst}) ]dt
\end{equation} and 

\begin{equation}
\label{eq:avg2} 
	A_{spiking} = \frac{1}{T}\int_0^T  \frac{1}{N_{I}} \sum_{i=1}^{N_I} \mathrm{sp}_i(t) dt .
\end{equation} 
Then, we explore the values of the ratio of bursting over spiking activity 
\begin{equation}
	\mathrm{ratio} = \frac{A_{bursting}}{A_{spiking}}
\end{equation}
as function of the coupling coefficient $\gamma$ and of the mean external input $\nu$ over the
parameter space $\mathcal{P}_1 = [0;\gamma_{max}] \times [0;\nu_{max}]$.

\subsection*{Quantification of oscillation power and frequency}\label{sec:fourier}

To quantify the frequency and the power of the oscillations in the neuronal activity, we perform a Fourier analysis
of the population activity which  we define as the sum of neuron spikes within a population, during the time step $dt$ 

\begin{equation}\label{eq:PA}
 r(t) = \frac{1}{dt} \frac{1}{N_I}\int_{t}^{t+dt} \sum_{i=1}^{N_I} \sum_{t_{ik} < t}  \delta(u-t_{ik}) du.
 \end{equation}

We compute a Discrete Time Fourier
Transform (DFT) and extract the power and the frequency of the most represented frequency in the Fourier
domain.
The formula defining the DFT is
\begin{equation}
	\hat{r}_k =  \sum_{n=0}^{N-1} r_n \exp{\left(-{i 2\pi k \frac{n}{N}} \right)}
	\qquad
	k = 0,\dots,N-1.
\end{equation}
where the ${r_n}$ sequence represents $N$ uniformly spaced time-samples of the population activities.
We measure the amplitude of the Fourier components $\hat{r}_k$ for $k= 1..N/2$ (because the
Fourier signal is symmetric from $N/2$ to $N$). We identify the maximal one,
its associated frequency $f_{max} = \frac{k}{N}$ and its power $P = (|\hat{r}_k|/N)^2$.

\subsection*{Downstream read-out neurons.}

To simulate the projection of a cortical layer onto another layer, we model
downstream read-out neurons with the same regular spiking neuron model as the first cortical layer.
The input $I_j$ received by each downstream neuron is the projected activity of all excitatory and inhibitory neurons of the first cortical layer, multiplied by the coefficients $W^{ERON}$ and  $W^{IRON}$ respectively:
\begin{equation}
	I_j(t) = W^{ERON} \sum_{i=1}^{N_E} \sum_{
				t_{ik}<t
			}
				\exp{\left(-\frac{t-t_{ik}}{\tau_{I_E}}\right)} 
				+  
				W^{IRON}\sum_{i=1}^{N_I} \sum_{
				t_{ik}<t
			}
				\exp{\left(-\frac{t-t_{ik}}{\tau_{I_I}}\right)}  .
\end{equation}

While giving the step current $I^{step}$, we introduce jitters so that the step current is not received by all neurons at the same time to avoid synchronization of the network due to the simultaneous strong common input. 
For each neuron, the time of current transition is drawn from a Gaussian distribution centered on the transition time and with variance 10 ms.

\subsection*{Cross-network synchronization.}
We investigate the role of gap junction coupling and its plasticity in synchronizing networks having different oscillation frequency preferences. We design a network consisting of two subnetworks having the same topology as described in \textit{Network} \ref{sec:network}: Each subnetworks has 800 excitatory neurons and 200 inhibitory neurons. 
There are all-to-all chemical synapses within each subnetworks (their strengths are reported in Table 1). There are no cross-network chemical synapses. The intra-network gap junctions are all-to-all. 
In addition, we vary the number of sparse cross-network gap junctions from 0 to 50. The gap junction strengths are initialized following a log-normal distribution as described in \textit{Network}. We take $\gamma =3$ to set the network in the AI regime and we take $\gamma = 5.5$ (which corresponds to the plasticity steady-state) to initialize the network in the SR regime.

One of the networks is called the Slow Network (SN) and we change the value of the membrane time constant of its inhibitory neurons $\tau_{v_S}$ from 17 ms to 55 ms. 
This decreases the neuron subthreshold resonance, which also lowers the frequency of its oscillation when it is in the synchronous regime. 
The second network has its neuron membrane time constant at 17 ms and is called the gamma-network because it oscillates at gamma frequency.
The simulations last 10 seconds, which is long enough for the gap junction coupling to reach its steady state when the gap junction are plastic.

To quantify the similarity between population activities from both subnetworks, we evaluate the Pearson's correlation coefficient between their population activities $r_{GN}$ and $r_{SN}$ from the gamma- and slow-network respectively. The firing rates, $r_{GN}$ and $r_{SN}$ are defined as in equation (\ref{eq:PA}).

For each subnetwork, we evaluate the frequency and power of their oscillations as described in the section \textit{Quantification of oscillation power and frequency}.
When the difference of oscillation frequency between both networks is less than 1 Hz, we measure the cross-correlation of their population activities $r_{GN}$ and $r_{SN}$

\begin{equation}
 (r_{GN}\star r_{SN})(\tau )\ {\stackrel {\mathrm {def} }{=}}\int _{-\infty }^{\infty }r_{GN}(t)\ r_{SN}(t+\tau )\,dt.
\end{equation} 

The phase difference is measured as the time delay relative to the oscillation period

\begin{equation}
 \Delta \phi = \frac {{\underset {t}{\operatorname {arg\,max} }}((r_{GN}\star r_{SN})(\tau)) } {T_{period}}.
\end{equation}

$\star$ is the convolution operator and $T_{period}$ is the oscillation period.

\subsection*{Information routing}

We investigate whether gap junction coupling and its plasticity play a role in routing information between networks.
We consider the same system as described in the previous section, with two subnetworks coupled with gap junctions, except here all the inhibitory neurons have the same membrane time constant $\tau_v$ = 17 ms (e.g. corresponding to resonance frequency at gamma). 
The first network, called the Input Network (IN) receives an input projected to its $N_{IN}$ neurons ($N_{IN}=1000$) by $N_{IN}$ weights drawn from a uniform distribution between 0 and 1. 
The second network is called the Output Network (ON, $N_{ON} = 1000$). 

To examine if there is successful transfer of information between both networks, we try to reconstruct the input signal from the ON's population activity $r_{ON}$.
First, we obtain the low-pass filtered population activity of ON, $r_{filt}$, with 
\begin{equation}\label{eq:p} 
	\tau_{r} \dot{r}_{filt}(t) = -r_{filt}(t) + r_{ON}(t),
\end{equation}
with $\tau_r$ = 3 ms.
Then we detect the rising and falling times of the filtered population activity by detecting when it crosses a  threshold $\theta_r$ = 2. 
This gives us rising times $ t^*_k$, when it crosses the threshold from below and falling times, when it crosses the threshold from above. 
We obtain the peak intervals $T_k$ by measuring the time difference between consecutive rising times $T_k =  t^*_{k+1} -  t^*_{k}$.

For Figure 5D, we plot $\overline{x}_k$, the mean values of the input signal $x$ between the rising times $ t^*_{k}$ and $ t^*_{k+1}$ as function of their corresponding peak intervals $T_k$
\begin{equation}
	\overline{x}_k(t) = \frac{1}{T_k} \int_{ t^*_k}^{ t^*_{k+1}} x(t) dt.
\end{equation}

We reconstruct the network input (Figure 5 E,H) by doing a linear interpolation of the inverse of those peak intervals $T_k$, so that the input signal and reconstructed input have the same length.

\begin{equation}
 \hat{x}(t) = \left(\frac{ 1/T_{k+1} - 1/T_{k}}{ t^*_{k+1} -  t^*_k}\right)(t-t^*_k) + \frac{1}{T_k}, \forall t \in [ t^*_k ;  t^*_{k+1}].
\end{equation}
Finally to estimate the quality of the reconstruction, we measure the Pearson's correlation coefficient (which is invariant by affine transformation) between the input and the reconstructed input. 

In order to test the robustness of the system we measure the quality of the reconstruction for an oscillatory input signal of which we vary the frequency $f$ (Figure 5F) and amplitude $A$ (figure 5G). 
\begin{equation}
	x(t) = A  [cos(2 \pi f t) + 1]
\end{equation}

Then we measure the routing of random signals $x(t) = \nu_{IN} + \sigma_{IN} \eta_{IN}$, where $\nu_{IN}$ is the signal mean, $\sigma_{IN}$ is the signal standard deviation, $\eta_{IN}$ is an Ornstein Uhlenbeck fluctuation with correlation time $\tau_x$ = 100 ms and unit variance. We build a dataset of 10 input signals and then we measure the Pearson's correlation coefficients between the input $x(t)$ and the reconstructed input $\hat{x}(t)$ for those 10 inputs respectively. 
For Figure 5I, we scale the log-normal distribution of the gap junction strength (see \textit{Network}) with $\gamma = 3$ to set the network in the asynchronous, with $\gamma = 5.5$ to set the gap junction near their plasticity fix point, and $\gamma = 8$ for a regime with strong oscillations.

To study the robustness of the information routing to gap junction deletion, we randomly delete an increasing number of gap junctions and measure the evolution of the Pearson's correlation between $x$ and $\hat{x}$. We also measure the change in the mean gap junction coupling, if there is plasticity, between the initialization (with $\gamma = 5.5$) and the steady-state (after 6 s of simulation).

All parameters are listed in Table 1 unless otherwise specified in a figure.


\subsection*{Parameters}
We list in Table 1 the parameters used for our simulations.

\begin{table}[htbp]  
\caption {} \label{tab:1}

\begin{tabular}[t]{l l}

	\begin{tabular}[t]{l}
	\textbf{Cortical Fast Spiking Interneurons}  \\
	\begin{tabularx}{.4\linewidth}{l r}
	$\tau_{I_I}$	&	10 ms	\\
	$\tau_v$	&	17 ms \\
	$\tau_v$ for SN (Fig. 3 and 4) 	&	[17-55] ms \\
    $\tau_u$	&	10 ms \\
	$R$ &	8 $\Omega$	\\
	$k_u$ &	10 $\Omega$	\\
	$v_{ra}$ & -75 mV	\\
	$v_{rb}$ &	-60 mV	\\
	$v_{rc}$	& -64 mV	\\
    $v_{resetFS}$	& -47 mV	\\
	$v_{threshFS}$	& 25 mV	\\
    $a$ & $1\Omega^{-1}$ \\
	$b$ & 50 pA	\\
    $k_{spikelet}$ &	40 \\
   \end{tabularx} \\ \\
 \end{tabular} 
 
 &
 
 \begin{tabular}[t]{l}
 \textbf{Cortical Regular Spiking Neurons} \\
   \begin{tabularx}{.4\linewidth}{l r}
	$\tau_{I_E}$	& 12 ms	\\
	$\tau_m$	& 40 ms \\
	$R_{m}$	& 0.6 $\Omega$	\\
	$v_{resetRS}$	&	-70 mV	\\
  	$v_{threshRS}$ &	0 mV \\
   \end{tabularx}\\ \\
   
   \textbf{Gap junction plasticity} \\
  	\begin{tabularx}{.4\linewidth}{l r}
	$\alpha_{LTD}$	& $1.569 . 10^{-5}$ $\Omega^{-1}s^{-1}$	\\
	$\alpha_{LTP}$	&	2.9 $\alpha_{LTD}$ \\
	$\theta_{burst}$	&	1.3\\ 
	$\tau_{b}$	&	8 ms \\
	\end{tabularx}\\ \\
	
	\textbf{Downstream read-out neuron}\\
	\begin{tabularx}{.4\linewidth}{l r}
	$T_{sim}$ & 10 s \\
	$N_{RON}$ & 200 \\
	$\nu_I$ 	&	20 pA 	\\ 
	$I^{step}$ 	&	250 pA	\\
	$W^{IE}$	&	-10000	\\ 
	$W^{ERON}$	&	1000	\\
	$W^{IRON}$	&	-1750	\\	
    \end{tabularx} \\ \\ 
  \end{tabular} \\ \\
	
  \begin{tabular}[t]{l}
  
   \textbf{Network} \\
   \begin{tabularx}{.4\linewidth}{l r}
    dt &	0.1ms	\\
	$N_I$	&	200	\\
	$N_E$	&	800 \\
	$W^{II}$	& $-80$	\\
	$W^{IE}$	& $-5000$	\\
	$W^{EE}$	& $500$	\\
	$W^{EI}$	& $300$	\\
	$\gamma$ (Fig. 2E, 2F, 3) & 5.5 \\
	$\gamma$ (Fig. 5 other than I) & 5.5 \\
	$\gamma$ (Fig. 4) for GN & 3 \\
	$\gamma$ (Fig. 4) for SN & 5.5 \\
	$\gamma_b$ (Fig. 1 and 2)	&	10	\\ 
	$\gamma_b$ (Fig. 3-5)	&	0	\\ 
    $\sigma_{gap}$ & 1 \\
    $\mu_{gap}$ & 1 \\
    $\sigma_I$	&	400 \\
	$\nu_I$ &	[0 pA; 300 pA]	\\ 
	$\tau_{noise}$ & 10 ms \\
	\end{tabularx} 
\end{tabular} &

\begin{tabular}[t]{l}
  
   \textbf{Information routing - Figure 5} \\
   \begin{tabularx}{.4\linewidth}{l r}
    $T_{sim}$ & 10 s \\
    	$\tau_{filt}$ & 3 ms \\
    	$\tau_{x}$ & 100 ms \\
    	$\mu_{IN}$ & 0.5\\
    	$\sigma_{IN}$ & 1/200 \\
    	$\nu_I$ 	&	200 pA 	\\ 
    	$\theta_{r}$ & 2 \\
    	$A$ (Fig. 5D) & [0-2000] pA \\
    	$A$ (Fig. 5G) & [0-10000] pA \\
    	$A$ (Fig. 5, all others) & 400 pA \\
    	$f$ (Fig. 5B,C,D,E) & 4 Hz \\
		
	\end{tabularx} 
\end{tabular}

\end{tabular}
 
\label{table:params1} 
\end{table}

\newpage
\bibliographystyle{elsarticle-num}
\bibliography{mendeley}

%
%

\newpage
\begin{figure}[H]
\begin{fullwidth}
	\begin{tabular}{ p{14cm}}
		\begin{tabular}{p{5cm} p{5cm} p{5cm}}
		
			\textbf{\large{A}} &  \textbf{\large{B}} & \textbf{\large{C}} \\
			\usetikzlibrary{matrix,chains,positioning,decorations.pathreplacing,arrows,arrows.meta}
\usetikzlibrary{matrix,positioning,calc}
\usetikzlibrary{decorations.pathmorphing}

\xdefinecolor{c1}{HTML}{4ED99C}
\xdefinecolor{Ecol}{HTML}{FF6868}
\xdefinecolor{Icol}{HTML}{3366cc}
\begin{tikzpicture}[
           > = Stealth, semithick, 
plain/.style = {draw=none, fill=none, yshift=11mm,
                text width=7ex,  align=center},
   ec/.style = {draw=none},
  net/.style = {
    matrix of nodes,
    nodes={circle, draw, semithick, minimum size=5.1mm, inner sep=0mm},
    nodes in empty cells,
  column sep = 3mm, 
     row sep = 5mm  
            },
]
\matrix[net] (m)
{
   E                   & |[ec]|                  & |[ec]|    &   E    \\
|[ec]|                &     I                    &  I          & |[ec]|            \\
   E                   & |[ec]|                  & |[ec]|      &    E       \\
};
\% inputs

\node[align=center, yshift=1.7em] (title) 
    at (current bounding box.north)
    {E-I recurrent network w/ GJs};

\tikzset{E/.style args={[#1]#2}{
    draw,
    circle,
    fill=Ecol,
    minimum size=5mm,
    label={[white,#1]center:#2}
    }
}
\tikzset{I/.style args={[#1]#2}{
    draw,
    circle,
    fill=Icol,
    minimum size=5mm,
    label={[white,#1]center:#2}
    }
}

\foreach \row in {1,3}
    \foreach \col in {1,4}
    		\node at (m-\row-\col) [E={[font=\large]E}] {};
    		
\node at (m-2-2) [I={[font=\large]I}] {};
\node at (m-2-3) [I={[font=\large]I}] {};

\path[-triangle 90 reversed]
(m-3-1) edge [bend left=-10,looseness=1] (m-3-4)
        edge [bend left=10,looseness=1] (m-1-1)
        edge [bend left=10,looseness=1] (m-2-2)
(m-1-1) edge [bend left=10,looseness=1] (m-1-4)
        edge [bend left=10,looseness=1] (m-3-1)
        edge [bend left=10,looseness=1] (m-2-2)
(m-1-4) edge [bend left=10,looseness=1] (m-3-4)
        edge [bend left=10,looseness=1] (m-1-1)
        edge [bend left=10,looseness=1] (m-2-3)
(m-3-4) edge [bend left=10,looseness=1] (m-1-4)
        edge [bend left=-10,looseness=1] (m-3-1)
        edge [bend left=10,looseness=1] (m-2-3);
        
\path[-*]
	(m-2-2) edge [bend left=30,looseness=1] (m-2-3)
	(m-2-3) edge [bend left=30,looseness=1] (m-2-2)
	(m-2-2) edge [bend left=10,looseness=1] (m-1-1)
			edge [bend left=10,looseness=1] (m-3-1)
	(m-2-3) edge [bend left=10,looseness=1] (m-1-4)
			edge [bend left=10,looseness=1] (m-3-4);

\path
	(m-2-2) edge [decoration = {zigzag,segment length = 2mm, amplitude = 0.5mm},decorate,ultra thick,c1] (m-2-3);

\end{tikzpicture}
			&
			\resizebox{5cm}{!}{\input{figures1/figure1-resonance.pgf}}

			&
			\resizebox{5cm}{!}{\input{figures1/figure1-d-coupling.pgf}}

		\end{tabular}
		\begin{tabular}{p{5cm} p{5cm} p{5cm}}
			\textbf{\large{D}} &  \textbf{\large{E}} & \textbf{\large{F}} \\
			\resizebox{5cm}{!}{
\begingroup%
\makeatletter%
\begin{pgfpicture}%
\pgfpathrectangle{\pgfpointorigin}{\pgfqpoint{4.884582in}{3.922941in}}%
\pgfusepath{use as bounding box, clip}%
\begin{pgfscope}%
\pgfsetbuttcap%
\pgfsetmiterjoin%
\pgfsetlinewidth{0.000000pt}%
\definecolor{currentstroke}{rgb}{0.000000,0.000000,0.000000}%
\pgfsetstrokecolor{currentstroke}%
\pgfsetstrokeopacity{0.000000}%
\pgfsetdash{}{0pt}%
\pgfpathmoveto{\pgfqpoint{0.000000in}{0.000000in}}%
\pgfpathlineto{\pgfqpoint{4.884582in}{0.000000in}}%
\pgfpathlineto{\pgfqpoint{4.884582in}{3.922941in}}%
\pgfpathlineto{\pgfqpoint{0.000000in}{3.922941in}}%
\pgfpathclose%
\pgfusepath{}%
\end{pgfscope}%
\begin{pgfscope}%
\pgfsetbuttcap%
\pgfsetmiterjoin%
\pgfsetlinewidth{0.000000pt}%
\definecolor{currentstroke}{rgb}{0.000000,0.000000,0.000000}%
\pgfsetstrokecolor{currentstroke}%
\pgfsetstrokeopacity{0.000000}%
\pgfsetdash{}{0pt}%
\pgfpathmoveto{\pgfqpoint{0.817312in}{0.821639in}}%
\pgfpathlineto{\pgfqpoint{3.890175in}{0.821639in}}%
\pgfpathlineto{\pgfqpoint{3.890175in}{3.507457in}}%
\pgfpathlineto{\pgfqpoint{0.817312in}{3.507457in}}%
\pgfpathclose%
\pgfusepath{}%
\end{pgfscope}%
\begin{pgfscope}%
\definecolor{textcolor}{rgb}{0.150000,0.150000,0.150000}%
\pgfsetstrokecolor{textcolor}%
\pgfsetfillcolor{textcolor}%
\pgftext[x=0.817312in,y=0.724417in,,top]{\color{textcolor}\fontsize{20.000000}{24.000000}\selectfont 0}%
\end{pgfscope}%
\begin{pgfscope}%
\definecolor{textcolor}{rgb}{0.150000,0.150000,0.150000}%
\pgfsetstrokecolor{textcolor}%
\pgfsetfillcolor{textcolor}%
\pgftext[x=2.353744in,y=0.724417in,,top]{\color{textcolor}\fontsize{20.000000}{24.000000}\selectfont 5}%
\end{pgfscope}%
\begin{pgfscope}%
\definecolor{textcolor}{rgb}{0.150000,0.150000,0.150000}%
\pgfsetstrokecolor{textcolor}%
\pgfsetfillcolor{textcolor}%
\pgftext[x=3.890175in,y=0.724417in,,top]{\color{textcolor}\fontsize{20.000000}{24.000000}\selectfont 10}%
\end{pgfscope}%
\begin{pgfscope}%
\definecolor{textcolor}{rgb}{0.150000,0.150000,0.150000}%
\pgfsetstrokecolor{textcolor}%
\pgfsetfillcolor{textcolor}%
\pgftext[x=2.353744in,y=0.400035in,,top]{\color{textcolor}\fontsize{22.000000}{26.400000}\selectfont Gap junction coupling, \(\displaystyle \gamma\)}%
\end{pgfscope}%
\begin{pgfscope}%
\definecolor{textcolor}{rgb}{0.150000,0.150000,0.150000}%
\pgfsetstrokecolor{textcolor}%
\pgfsetfillcolor{textcolor}%
\pgftext[x=0.662310in,y=0.750431in,left,base,rotate=90.000000]{\color{textcolor}\fontsize{20.000000}{24.000000}\selectfont 0}%
\end{pgfscope}%
\begin{pgfscope}%
\definecolor{textcolor}{rgb}{0.150000,0.150000,0.150000}%
\pgfsetstrokecolor{textcolor}%
\pgfsetfillcolor{textcolor}%
\pgftext[x=0.662310in,y=1.739879in,left,base,rotate=90.000000]{\color{textcolor}\fontsize{20.000000}{24.000000}\selectfont 150}%
\end{pgfscope}%
\begin{pgfscope}%
\definecolor{textcolor}{rgb}{0.150000,0.150000,0.150000}%
\pgfsetstrokecolor{textcolor}%
\pgfsetfillcolor{textcolor}%
\pgftext[x=0.662310in,y=3.082788in,left,base,rotate=90.000000]{\color{textcolor}\fontsize{20.000000}{24.000000}\selectfont 300}%
\end{pgfscope}%
\begin{pgfscope}%
\definecolor{textcolor}{rgb}{0.150000,0.150000,0.150000}%
\pgfsetstrokecolor{textcolor}%
\pgfsetfillcolor{textcolor}%
\pgftext[x=0.395708in,y=2.164548in,,bottom,rotate=90.000000]{\color{textcolor}\fontsize{22.000000}{26.400000}\selectfont Network drive, \(\displaystyle \nu\)}%
\end{pgfscope}%
\begin{pgfscope}%
\pgfsys@transformshift{0.819444in}{0.825718in}%
\pgftext[left,bottom]{\pgfimage[interpolate=true,width=3.069444in,height=2.694444in]{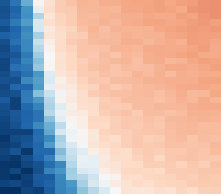}}%
\end{pgfscope}%
\begin{pgfscope}%
\definecolor{textcolor}{rgb}{1.0,1.0,1.0}%
\pgfsetstrokecolor{textcolor}%
\pgfsetfillcolor{textcolor}%
\pgftext[x=0.9in,y=1.716912in,left,base]{\color{textcolor}\fontsize{30.000000}{36.000000}\selectfont \ AI}%
\end{pgfscope}%
\begin{pgfscope}%
\definecolor{textcolor}{rgb}{0.0,0.0,0.0}%
\pgfsetstrokecolor{textcolor}%
\pgfsetfillcolor{textcolor}%
\pgftext[x=2.408165in,y=1.716912in,left,base]{\color{textcolor}\fontsize{30.000000}{36.000000}\selectfont \ SR}%
\end{pgfscope}%
\begin{pgfscope}%
\pgftext[x=2.353744in,y=3.590790in,,base]{\fontsize{22.000000}{26.400000}\selectfont Oscillation Power}%
\end{pgfscope}%
\begin{pgfscope}%
\pgfpathrectangle{\pgfqpoint{4.082229in}{0.821639in}}{\pgfqpoint{0.134291in}{2.685818in}} %
\pgfusepath{clip}%
\pgfsetbuttcap%
\pgfsetmiterjoin%
\definecolor{currentfill}{rgb}{1.000000,1.000000,1.000000}%
\pgfsetfillcolor{currentfill}%
\pgfsetlinewidth{0.010037pt}%
\definecolor{currentstroke}{rgb}{1.000000,1.000000,1.000000}%
\pgfsetstrokecolor{currentstroke}%
\pgfsetdash{}{0pt}%
\pgfpathmoveto{\pgfqpoint{4.082229in}{0.821639in}}%
\pgfpathlineto{\pgfqpoint{4.082229in}{0.832131in}}%
\pgfpathlineto{\pgfqpoint{4.082229in}{3.496965in}}%
\pgfpathlineto{\pgfqpoint{4.082229in}{3.507457in}}%
\pgfpathlineto{\pgfqpoint{4.216520in}{3.507457in}}%
\pgfpathlineto{\pgfqpoint{4.216520in}{3.496965in}}%
\pgfpathlineto{\pgfqpoint{4.216520in}{0.832131in}}%
\pgfpathlineto{\pgfqpoint{4.216520in}{0.821639in}}%
\pgfpathclose%
\pgfusepath{stroke,fill}%
\end{pgfscope}%
\begin{pgfscope}%
\definecolor{textcolor}{rgb}{0.150000,0.150000,0.150000}%
\pgfsetstrokecolor{textcolor}%
\pgfsetfillcolor{textcolor}%
\pgftext[x=4.313742in,y=1.073298in,left,base]{\color{textcolor}\fontsize{20.000000}{24.000000}\selectfont \(\displaystyle -20\)}%
\end{pgfscope}%
\begin{pgfscope}%
\definecolor{textcolor}{rgb}{0.150000,0.150000,0.150000}%
\pgfsetstrokecolor{textcolor}%
\pgfsetfillcolor{textcolor}%
\pgftext[x=4.313742in,y=1.566161in,left,base]{\color{textcolor}\fontsize{20.000000}{24.000000}\selectfont \(\displaystyle -10\)}%
\end{pgfscope}%
\begin{pgfscope}%
\definecolor{textcolor}{rgb}{0.150000,0.150000,0.150000}%
\pgfsetstrokecolor{textcolor}%
\pgfsetfillcolor{textcolor}%
\pgftext[x=4.313742in,y=2.059025in,left,base]{\color{textcolor}\fontsize{20.000000}{24.000000}\selectfont \(\displaystyle 0\)}%
\end{pgfscope}%
\begin{pgfscope}%
\definecolor{textcolor}{rgb}{0.150000,0.150000,0.150000}%
\pgfsetstrokecolor{textcolor}%
\pgfsetfillcolor{textcolor}%
\pgftext[x=4.313742in,y=2.551888in,left,base]{\color{textcolor}\fontsize{20.000000}{24.000000}\selectfont \(\displaystyle 10\)}%
\end{pgfscope}%
\begin{pgfscope}%
\definecolor{textcolor}{rgb}{0.150000,0.150000,0.150000}%
\pgfsetstrokecolor{textcolor}%
\pgfsetfillcolor{textcolor}%
\pgftext[x=4.313742in,y=3.044752in,left,base]{\color{textcolor}\fontsize{20.000000}{24.000000}\selectfont \(\displaystyle 20\)}%
\end{pgfscope}%
\begin{pgfscope}%
\pgfsys@transformshift{4.083333in}{0.825718in}%
\pgftext[left,bottom]{\pgfimage[interpolate=true,width=0.138889in,height=2.694444in]{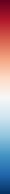}}%
\end{pgfscope}%
\begin{pgfscope}%
\pgfsetbuttcap%
\pgfsetmiterjoin%
\pgfsetlinewidth{0.000000pt}%
\definecolor{currentstroke}{rgb}{1.000000,1.000000,1.000000}%
\pgfsetstrokecolor{currentstroke}%
\pgfsetdash{}{0pt}%
\pgfpathmoveto{\pgfqpoint{4.082229in}{0.821639in}}%
\pgfpathlineto{\pgfqpoint{4.082229in}{0.832131in}}%
\pgfpathlineto{\pgfqpoint{4.082229in}{3.496965in}}%
\pgfpathlineto{\pgfqpoint{4.082229in}{3.507457in}}%
\pgfpathlineto{\pgfqpoint{4.216520in}{3.507457in}}%
\pgfpathlineto{\pgfqpoint{4.216520in}{3.496965in}}%
\pgfpathlineto{\pgfqpoint{4.216520in}{0.832131in}}%
\pgfpathlineto{\pgfqpoint{4.216520in}{0.821639in}}%
\pgfpathclose%
\pgfusepath{}%
\end{pgfscope}%
\end{pgfpicture}%
\makeatother%
\endgroup%
}
			&
			\resizebox{5cm}{!}{
\begingroup%
\makeatletter%
\begin{pgfpicture}%
\pgfpathrectangle{\pgfpointorigin}{\pgfqpoint{4.784178in}{3.922941in}}%
\pgfusepath{use as bounding box, clip}%
\begin{pgfscope}%
\pgfsetbuttcap%
\pgfsetmiterjoin%
\pgfsetlinewidth{0.000000pt}%
\definecolor{currentstroke}{rgb}{0.000000,0.000000,0.000000}%
\pgfsetstrokecolor{currentstroke}%
\pgfsetstrokeopacity{0.000000}%
\pgfsetdash{}{0pt}%
\pgfpathmoveto{\pgfqpoint{0.000000in}{0.000000in}}%
\pgfpathlineto{\pgfqpoint{4.784178in}{0.000000in}}%
\pgfpathlineto{\pgfqpoint{4.784178in}{3.922941in}}%
\pgfpathlineto{\pgfqpoint{0.000000in}{3.922941in}}%
\pgfpathclose%
\pgfusepath{}%
\end{pgfscope}%
\begin{pgfscope}%
\pgfsetbuttcap%
\pgfsetmiterjoin%
\pgfsetlinewidth{0.000000pt}%
\definecolor{currentstroke}{rgb}{0.000000,0.000000,0.000000}%
\pgfsetstrokecolor{currentstroke}%
\pgfsetstrokeopacity{0.000000}%
\pgfsetdash{}{0pt}%
\pgfpathmoveto{\pgfqpoint{0.817312in}{0.821639in}}%
\pgfpathlineto{\pgfqpoint{3.999018in}{0.821639in}}%
\pgfpathlineto{\pgfqpoint{3.999018in}{3.507457in}}%
\pgfpathlineto{\pgfqpoint{0.817312in}{3.507457in}}%
\pgfpathclose%
\pgfusepath{}%
\end{pgfscope}%
\begin{pgfscope}%
\definecolor{textcolor}{rgb}{0.150000,0.150000,0.150000}%
\pgfsetstrokecolor{textcolor}%
\pgfsetfillcolor{textcolor}%
\pgftext[x=0.817312in,y=0.724417in,,top]{\color{textcolor}\fontsize{20.000000}{24.000000}\selectfont 0}%
\end{pgfscope}%
\begin{pgfscope}%
\definecolor{textcolor}{rgb}{0.150000,0.150000,0.150000}%
\pgfsetstrokecolor{textcolor}%
\pgfsetfillcolor{textcolor}%
\pgftext[x=2.408165in,y=0.724417in,,top]{\color{textcolor}\fontsize{20.000000}{24.000000}\selectfont 5}%
\end{pgfscope}%
\begin{pgfscope}%
\definecolor{textcolor}{rgb}{0.150000,0.150000,0.150000}%
\pgfsetstrokecolor{textcolor}%
\pgfsetfillcolor{textcolor}%
\pgftext[x=3.999018in,y=0.724417in,,top]{\color{textcolor}\fontsize{20.000000}{24.000000}\selectfont 10}%
\end{pgfscope}%
\begin{pgfscope}%
\definecolor{textcolor}{rgb}{0.150000,0.150000,0.150000}%
\pgfsetstrokecolor{textcolor}%
\pgfsetfillcolor{textcolor}%
\pgftext[x=2.408165in,y=0.400035in,,top]{\color{textcolor}\fontsize{22.000000}{26.400000}\selectfont Gap junction coupling, \(\displaystyle \gamma\)}%
\end{pgfscope}%
\begin{pgfscope}%
\definecolor{textcolor}{rgb}{0.150000,0.150000,0.150000}%
\pgfsetstrokecolor{textcolor}%
\pgfsetfillcolor{textcolor}%
\pgftext[x=0.662310in,y=0.750431in,left,base,rotate=90.000000]{\color{textcolor}\fontsize{20.000000}{24.000000}\selectfont 0}%
\end{pgfscope}%
\begin{pgfscope}%
\definecolor{textcolor}{rgb}{0.150000,0.150000,0.150000}%
\pgfsetstrokecolor{textcolor}%
\pgfsetfillcolor{textcolor}%
\pgftext[x=0.662310in,y=1.739879in,left,base,rotate=90.000000]{\color{textcolor}\fontsize{20.000000}{24.000000}\selectfont 150}%
\end{pgfscope}%
\begin{pgfscope}%
\definecolor{textcolor}{rgb}{0.150000,0.150000,0.150000}%
\pgfsetstrokecolor{textcolor}%
\pgfsetfillcolor{textcolor}%
\pgftext[x=0.662310in,y=3.082788in,left,base,rotate=90.000000]{\color{textcolor}\fontsize{20.000000}{24.000000}\selectfont 300}%
\end{pgfscope}%
\begin{pgfscope}%
\definecolor{textcolor}{rgb}{0.150000,0.150000,0.150000}%
\pgfsetstrokecolor{textcolor}%
\pgfsetfillcolor{textcolor}%
\pgftext[x=0.395708in,y=2.164548in,,bottom,rotate=90.000000]{\color{textcolor}\fontsize{22.000000}{26.400000}\selectfont Network drive, \(\displaystyle \nu\)}%
\end{pgfscope}%
\begin{pgfscope}%
\pgfsys@transformshift{1.291667in}{0.825718in}%
\pgftext[left,bottom]{\pgfimage[interpolate=true,width=2.708333in,height=2.694444in]{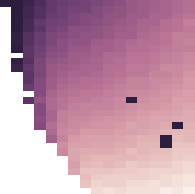}}%
\end{pgfscope}%
\begin{pgfscope}%
\definecolor{textcolor}{rgb}{0.0,0.0,0.0}%
\pgfsetstrokecolor{textcolor}%
\pgfsetfillcolor{textcolor}%
\pgftext[x=0.9in,y=1.716912in,left,base]{\color{textcolor}\fontsize{30.000000}{36.000000}\selectfont \ AI}%
\end{pgfscope}%
\begin{pgfscope}%
\definecolor{textcolor}{rgb}{1.000000,1.0,1.0}%
\pgfsetstrokecolor{textcolor}%
\pgfsetfillcolor{textcolor}%
\pgftext[x=2.408165in,y=1.716912in,left,base]{\color{textcolor}\fontsize{30.000000}{36.000000}\selectfont \ SR}%
\end{pgfscope}%
\begin{pgfscope}%
\pgftext[x=2.408165in,y=3.590790in,,base]{\fontsize{22.000000}{26.400000}\selectfont Oscillation Frequency}%
\end{pgfscope}%
\begin{pgfscope}%
\pgfpathrectangle{\pgfqpoint{4.197875in}{0.821639in}}{\pgfqpoint{0.134291in}{2.685818in}} %
\pgfusepath{clip}%
\pgfsetbuttcap%
\pgfsetmiterjoin%
\definecolor{currentfill}{rgb}{1.000000,1.000000,1.000000}%
\pgfsetfillcolor{currentfill}%
\pgfsetlinewidth{0.010037pt}%
\definecolor{currentstroke}{rgb}{1.000000,1.000000,1.000000}%
\pgfsetstrokecolor{currentstroke}%
\pgfsetdash{}{0pt}%
\pgfpathmoveto{\pgfqpoint{4.197875in}{0.821639in}}%
\pgfpathlineto{\pgfqpoint{4.197875in}{0.832131in}}%
\pgfpathlineto{\pgfqpoint{4.197875in}{3.496965in}}%
\pgfpathlineto{\pgfqpoint{4.197875in}{3.507457in}}%
\pgfpathlineto{\pgfqpoint{4.332165in}{3.507457in}}%
\pgfpathlineto{\pgfqpoint{4.332165in}{3.496965in}}%
\pgfpathlineto{\pgfqpoint{4.332165in}{0.832131in}}%
\pgfpathlineto{\pgfqpoint{4.332165in}{0.821639in}}%
\pgfpathclose%
\pgfusepath{stroke,fill}%
\end{pgfscope}%
\begin{pgfscope}%
\definecolor{textcolor}{rgb}{0.150000,0.150000,0.150000}%
\pgfsetstrokecolor{textcolor}%
\pgfsetfillcolor{textcolor}%
\pgftext[x=4.429388in,y=0.959682in,left,base]{\color{textcolor}\fontsize{20.000000}{24.000000}\selectfont \(\displaystyle 40\)}%
\end{pgfscope}%
\begin{pgfscope}%
\definecolor{textcolor}{rgb}{0.150000,0.150000,0.150000}%
\pgfsetstrokecolor{textcolor}%
\pgfsetfillcolor{textcolor}%
\pgftext[x=4.429388in,y=1.448132in,left,base]{\color{textcolor}\fontsize{20.000000}{24.000000}\selectfont \(\displaystyle 48\)}%
\end{pgfscope}%
\begin{pgfscope}%
\definecolor{textcolor}{rgb}{0.150000,0.150000,0.150000}%
\pgfsetstrokecolor{textcolor}%
\pgfsetfillcolor{textcolor}%
\pgftext[x=4.429388in,y=1.936583in,left,base]{\color{textcolor}\fontsize{20.000000}{24.000000}\selectfont \(\displaystyle 56\)}%
\end{pgfscope}%
\begin{pgfscope}%
\definecolor{textcolor}{rgb}{0.150000,0.150000,0.150000}%
\pgfsetstrokecolor{textcolor}%
\pgfsetfillcolor{textcolor}%
\pgftext[x=4.429388in,y=2.425033in,left,base]{\color{textcolor}\fontsize{20.000000}{24.000000}\selectfont \(\displaystyle 64\)}%
\end{pgfscope}%
\begin{pgfscope}%
\definecolor{textcolor}{rgb}{0.150000,0.150000,0.150000}%
\pgfsetstrokecolor{textcolor}%
\pgfsetfillcolor{textcolor}%
\pgftext[x=4.429388in,y=2.913483in,left,base]{\color{textcolor}\fontsize{20.000000}{24.000000}\selectfont \(\displaystyle 72\)}%
\end{pgfscope}%
\begin{pgfscope}%
\definecolor{textcolor}{rgb}{0.150000,0.150000,0.150000}%
\pgfsetstrokecolor{textcolor}%
\pgfsetfillcolor{textcolor}%
\pgftext[x=4.429388in,y=3.401934in,left,base]{\color{textcolor}\fontsize{20.000000}{24.000000}\selectfont \(\displaystyle 80\)}%
\end{pgfscope}%
\begin{pgfscope}%
\pgfsys@transformshift{4.194444in}{0.825718in}%
\pgftext[left,bottom]{\pgfimage[interpolate=true,width=0.138889in,height=2.694444in]{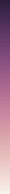}}%
\end{pgfscope}%
\begin{pgfscope}%
\pgfsetbuttcap%
\pgfsetmiterjoin%
\pgfsetlinewidth{0.000000pt}%
\definecolor{currentstroke}{rgb}{1.000000,1.000000,1.000000}%
\pgfsetstrokecolor{currentstroke}%
\pgfsetdash{}{0pt}%
\pgfpathmoveto{\pgfqpoint{4.197875in}{0.821639in}}%
\pgfpathlineto{\pgfqpoint{4.197875in}{0.832131in}}%
\pgfpathlineto{\pgfqpoint{4.197875in}{3.496965in}}%
\pgfpathlineto{\pgfqpoint{4.197875in}{3.507457in}}%
\pgfpathlineto{\pgfqpoint{4.332165in}{3.507457in}}%
\pgfpathlineto{\pgfqpoint{4.332165in}{3.496965in}}%
\pgfpathlineto{\pgfqpoint{4.332165in}{0.832131in}}%
\pgfpathlineto{\pgfqpoint{4.332165in}{0.821639in}}%
\pgfpathclose%
\pgfusepath{}%
\end{pgfscope}%
\end{pgfpicture}%
\makeatother%
\endgroup%
}
			&
			\resizebox{5cm}{!}{\input{figures1/figure1-suppl-frequency-histogram.pgf}}
			
		\end{tabular}
		\begin{tabular}{p{5cm} p{5cm} p{5cm}}
			\textbf{\large{G}} &  \textbf{\large{H}} & \textbf{\large{I}} \\
            \resizebox{5cm}{!}{
\begingroup%
\makeatletter%
\begin{pgfpicture}%
\pgfpathrectangle{\pgfpointorigin}{\pgfqpoint{4.901087in}{3.902893in}}%
\pgfusepath{use as bounding box, clip}%
\begin{pgfscope}%
\pgfsetbuttcap%
\pgfsetmiterjoin%
\pgfsetlinewidth{0.000000pt}%
\definecolor{currentstroke}{rgb}{0.000000,0.000000,0.000000}%
\pgfsetstrokecolor{currentstroke}%
\pgfsetstrokeopacity{0.000000}%
\pgfsetdash{}{0pt}%
\pgfpathmoveto{\pgfqpoint{0.000000in}{0.000000in}}%
\pgfpathlineto{\pgfqpoint{4.901087in}{0.000000in}}%
\pgfpathlineto{\pgfqpoint{4.901087in}{3.902893in}}%
\pgfpathlineto{\pgfqpoint{0.000000in}{3.902893in}}%
\pgfpathclose%
\pgfusepath{}%
\end{pgfscope}%
\begin{pgfscope}%
\pgftext[x=2.353744in,y=3.590790in,,base]{\fontsize{22.000000}{26.400000}\selectfont Bursting/Spiking Ratio}%
\end{pgfscope}%
\begin{pgfscope}%
\pgfsetbuttcap%
\pgfsetmiterjoin%
\pgfsetlinewidth{0.000000pt}%
\definecolor{currentstroke}{rgb}{0.000000,0.000000,0.000000}%
\pgfsetstrokecolor{currentstroke}%
\pgfsetstrokeopacity{0.000000}%
\pgfsetdash{}{0pt}%
\pgfpathmoveto{\pgfqpoint{0.817312in}{0.821639in}}%
\pgfpathlineto{\pgfqpoint{3.799083in}{0.821639in}}%
\pgfpathlineto{\pgfqpoint{3.799083in}{3.487409in}}%
\pgfpathlineto{\pgfqpoint{0.817312in}{3.487409in}}%
\pgfpathclose%
\pgfusepath{}%
\end{pgfscope}%
\begin{pgfscope}%
\definecolor{textcolor}{rgb}{0.150000,0.150000,0.150000}%
\pgfsetstrokecolor{textcolor}%
\pgfsetfillcolor{textcolor}%
\pgftext[x=0.817312in,y=0.724417in,,top]{\color{textcolor}\fontsize{20.000000}{24.000000}\selectfont 0}%
\end{pgfscope}%
\begin{pgfscope}%
\definecolor{textcolor}{rgb}{0.150000,0.150000,0.150000}%
\pgfsetstrokecolor{textcolor}%
\pgfsetfillcolor{textcolor}%
\pgftext[x=2.308197in,y=0.724417in,,top]{\color{textcolor}\fontsize{20.000000}{24.000000}\selectfont 5}%
\end{pgfscope}%
\begin{pgfscope}%
\definecolor{textcolor}{rgb}{0.150000,0.150000,0.150000}%
\pgfsetstrokecolor{textcolor}%
\pgfsetfillcolor{textcolor}%
\pgftext[x=3.799083in,y=0.724417in,,top]{\color{textcolor}\fontsize{20.000000}{24.000000}\selectfont 10}%
\end{pgfscope}%
\begin{pgfscope}%
\definecolor{textcolor}{rgb}{0.150000,0.150000,0.150000}%
\pgfsetstrokecolor{textcolor}%
\pgfsetfillcolor{textcolor}%
\pgftext[x=2.308197in,y=0.400035in,,top]{\color{textcolor}\fontsize{22.000000}{26.400000}\selectfont Gap junction coupling, \(\displaystyle \gamma\)}%
\end{pgfscope}%
\begin{pgfscope}%
\definecolor{textcolor}{rgb}{0.150000,0.150000,0.150000}%
\pgfsetstrokecolor{textcolor}%
\pgfsetfillcolor{textcolor}%
\pgftext[x=0.662310in,y=0.750431in,left,base,rotate=90.000000]{\color{textcolor}\fontsize{20.000000}{24.000000}\selectfont 0}%
\end{pgfscope}%
\begin{pgfscope}%
\definecolor{textcolor}{rgb}{0.150000,0.150000,0.150000}%
\pgfsetstrokecolor{textcolor}%
\pgfsetfillcolor{textcolor}%
\pgftext[x=0.662310in,y=1.729855in,left,base,rotate=90.000000]{\color{textcolor}\fontsize{20.000000}{24.000000}\selectfont 150}%
\end{pgfscope}%
\begin{pgfscope}%
\definecolor{textcolor}{rgb}{0.150000,0.150000,0.150000}%
\pgfsetstrokecolor{textcolor}%
\pgfsetfillcolor{textcolor}%
\pgftext[x=0.662310in,y=3.062740in,left,base,rotate=90.000000]{\color{textcolor}\fontsize{20.000000}{24.000000}\selectfont 300}%
\end{pgfscope}%
\begin{pgfscope}%
\definecolor{textcolor}{rgb}{0.150000,0.150000,0.150000}%
\pgfsetstrokecolor{textcolor}%
\pgfsetfillcolor{textcolor}%
\pgftext[x=0.395708in,y=2.154524in,,bottom,rotate=90.000000]{\color{textcolor}\fontsize{22.000000}{26.400000}\selectfont Network drive, \(\displaystyle \nu\)}%
\end{pgfscope}%
\begin{pgfscope}%
\pgfsys@transformshift{0.819444in}{0.819560in}%
\pgftext[left,bottom]{\pgfimage[interpolate=true,width=2.986111in,height=2.666667in]{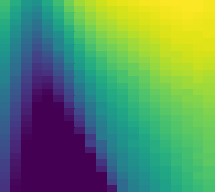}}%
\end{pgfscope}%
\begin{pgfscope}%
\definecolor{textcolor}{rgb}{1.0,1.0,1.0}%
\pgfsetstrokecolor{textcolor}%
\pgfsetfillcolor{textcolor}%
\pgftext[x=0.9in,y=1.716912in,left,base]{\color{textcolor}\fontsize{30.000000}{36.000000}\selectfont \ AI}%
\end{pgfscope}%
\begin{pgfscope}%
\definecolor{textcolor}{rgb}{0.0,0.0,0.0}%
\pgfsetstrokecolor{textcolor}%
\pgfsetfillcolor{textcolor}%
\pgftext[x=2.408165in,y=1.716912in,left,base]{\color{textcolor}\fontsize{30.000000}{36.000000}\selectfont \ SR}%
\end{pgfscope}%
\begin{pgfscope}%
\pgfpathrectangle{\pgfqpoint{3.985443in}{0.821639in}}{\pgfqpoint{0.133289in}{2.665770in}} %
\pgfusepath{clip}%
\pgfsetbuttcap%
\pgfsetmiterjoin%
\definecolor{currentfill}{rgb}{1.000000,1.000000,1.000000}%
\pgfsetfillcolor{currentfill}%
\pgfsetlinewidth{0.010037pt}%
\definecolor{currentstroke}{rgb}{1.000000,1.000000,1.000000}%
\pgfsetstrokecolor{currentstroke}%
\pgfsetdash{}{0pt}%
\pgfpathmoveto{\pgfqpoint{3.985443in}{0.821639in}}%
\pgfpathlineto{\pgfqpoint{3.985443in}{0.832052in}}%
\pgfpathlineto{\pgfqpoint{3.985443in}{3.476996in}}%
\pgfpathlineto{\pgfqpoint{3.985443in}{3.487409in}}%
\pgfpathlineto{\pgfqpoint{4.118732in}{3.487409in}}%
\pgfpathlineto{\pgfqpoint{4.118732in}{3.476996in}}%
\pgfpathlineto{\pgfqpoint{4.118732in}{0.832052in}}%
\pgfpathlineto{\pgfqpoint{4.118732in}{0.821639in}}%
\pgfpathclose%
\pgfusepath{stroke,fill}%
\end{pgfscope}%
\begin{pgfscope}%
\definecolor{textcolor}{rgb}{0.150000,0.150000,0.150000}%
\pgfsetstrokecolor{textcolor}%
\pgfsetfillcolor{textcolor}%
\pgftext[x=4.215954in,y=0.716116in,left,base]{\color{textcolor}\fontsize{20.000000}{24.000000}\selectfont \(\displaystyle 0.008\)}%
\end{pgfscope}%
\begin{pgfscope}%
\definecolor{textcolor}{rgb}{0.150000,0.150000,0.150000}%
\pgfsetstrokecolor{textcolor}%
\pgfsetfillcolor{textcolor}%
\pgftext[x=4.215954in,y=1.228763in,left,base]{\color{textcolor}\fontsize{20.000000}{24.000000}\selectfont \(\displaystyle 0.012\)}%
\end{pgfscope}%
\begin{pgfscope}%
\definecolor{textcolor}{rgb}{0.150000,0.150000,0.150000}%
\pgfsetstrokecolor{textcolor}%
\pgfsetfillcolor{textcolor}%
\pgftext[x=4.215954in,y=1.741410in,left,base]{\color{textcolor}\fontsize{20.000000}{24.000000}\selectfont \(\displaystyle 0.016\)}%
\end{pgfscope}%
\begin{pgfscope}%
\definecolor{textcolor}{rgb}{0.150000,0.150000,0.150000}%
\pgfsetstrokecolor{textcolor}%
\pgfsetfillcolor{textcolor}%
\pgftext[x=4.215954in,y=2.254057in,left,base]{\color{textcolor}\fontsize{20.000000}{24.000000}\selectfont \(\displaystyle 0.020\)}%
\end{pgfscope}%
\begin{pgfscope}%
\definecolor{textcolor}{rgb}{0.150000,0.150000,0.150000}%
\pgfsetstrokecolor{textcolor}%
\pgfsetfillcolor{textcolor}%
\pgftext[x=4.215954in,y=2.766704in,left,base]{\color{textcolor}\fontsize{20.000000}{24.000000}\selectfont \(\displaystyle 0.024\)}%
\end{pgfscope}%
\begin{pgfscope}%
\definecolor{textcolor}{rgb}{0.150000,0.150000,0.150000}%
\pgfsetstrokecolor{textcolor}%
\pgfsetfillcolor{textcolor}%
\pgftext[x=4.215954in,y=3.279351in,left,base]{\color{textcolor}\fontsize{20.000000}{24.000000}\selectfont \(\displaystyle 0.028\)}%
\end{pgfscope}%
\begin{pgfscope}%
\pgfsys@transformshift{3.986111in}{0.819560in}%
\pgftext[left,bottom]{\pgfimage[interpolate=true,width=0.138889in,height=2.666667in]{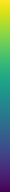}}%
\end{pgfscope}%
\begin{pgfscope}%
\pgfsetbuttcap%
\pgfsetmiterjoin%
\pgfsetlinewidth{0.000000pt}%
\definecolor{currentstroke}{rgb}{1.000000,1.000000,1.000000}%
\pgfsetstrokecolor{currentstroke}%
\pgfsetdash{}{0pt}%
\pgfpathmoveto{\pgfqpoint{3.985443in}{0.821639in}}%
\pgfpathlineto{\pgfqpoint{3.985443in}{0.832052in}}%
\pgfpathlineto{\pgfqpoint{3.985443in}{3.476996in}}%
\pgfpathlineto{\pgfqpoint{3.985443in}{3.487409in}}%
\pgfpathlineto{\pgfqpoint{4.118732in}{3.487409in}}%
\pgfpathlineto{\pgfqpoint{4.118732in}{3.476996in}}%
\pgfpathlineto{\pgfqpoint{4.118732in}{0.832052in}}%
\pgfpathlineto{\pgfqpoint{4.118732in}{0.821639in}}%
\pgfpathclose%
\pgfusepath{}%
\end{pgfscope}%
\end{pgfpicture}%
\makeatother%
\endgroup%
}
            &
			\resizebox{5cm}{!}{\input{figures1/figure1-f-raster.pgf}}	
			&
			\resizebox{5cm}{!}{\input{figures1/figure1-voltage-traces-inh.pgf}}	

		\end{tabular}

	\end{tabular}
	\caption{ {\bf Network synchrony depends on gap junction strength.} 
	\label{fig1}}
	(\textbf{A}) The network consists of excitatory (E) and inhibitory (I) neurons. The neurons are coupled in an all-to-all fashion with chemical synapses. The inhibitory neurons are also connected by gap junctions (jagged green line). 
(\textbf{B}) Voltage response of one single excitatory (red line) / inhibitory (blue line) neuron to a subthreshold oscillatory input current (see Methods). 
Excitatory neurons act as low-pass filters, whereas the inhibitory neurons show a resonance frequency in the gamma range. 
This resonance is in agreement with the network wide response observed by Cardin et al. 2009, when FS neurons are stimulated in the gamma range (black line, figure redrawn from [32] figure 3d).
(\textbf{C}) Simulation of a pair of electrically coupled neurons N1 and N2, where N1 is voltage-clamped (red) such that it is hyperpolarized (light blue) and the potential of N2 is measured for different value of gap junction strength ($\gamma = 3$ and $\gamma = 5$). 
(\textbf{D}) Power of the main frequency component in the Fourier domain of the population activity (PA) of inhibitory neurons. 
The blue area denotes the lack of oscillations \textbf{AI} whereas the red area \textbf{SR} shows periodic oscillations in the spiking activity of inhibitory neurons. 
(\textbf{E}) Oscillation frequency of the network activity. The white area represents a region where the network is not oscillating and has no oscillation frequency.
(\textbf{F}) Histogram of the oscillation frequency of population spiking activity. The values are contained in the $\gamma$ range, from 30 to 60Hz.
(\textbf{G}) Ratio of bursting $A_{bursting}$ over spiking $A_{spiking}$ activity, averaged over 2 seconds. Bursting activity prevails in the light region and sparse firing dominates in the dark region. For the following figures 1H and 1I, 100 ms of data is represented.
(\textbf{H}) Raster plots of 100 FS neurons (blue) and 100 pyramidal neurons (red) for two values of the gap junction coupling, where dots represents spiking times and each line represents a neuron (note that the network E/I proportion is actually 80\%/20\%). Top raster plot shows asynchronous activity for low gap junction coupling and bottom raster plot shows synchronous activity in inhibitory and excitatory neuron populations, for strong gap junction coupling.
(\textbf{I}) Membrane voltage traces of individual inhibitory neurons (dark blue) and population average (light blue, down-shifted) for different values of the gap junction coupling. Bursts appear for strong gap junction coupling on the peaks of the membrane voltage oscillations. 
\end{fullwidth}
\end{figure}

%
%

\newpage
\begin{figure}[H]
\begin{fullwidth}
	\begin{tabular}{ p{14cm}}
	
		\begin{tabular}{p{5cm} p{5cm} p{5cm}}
			\textbf{\large{A}} &  \textbf{\large{B}} & \textbf{\large{C}} \\
			
			\def\svgwidth{5cm}
			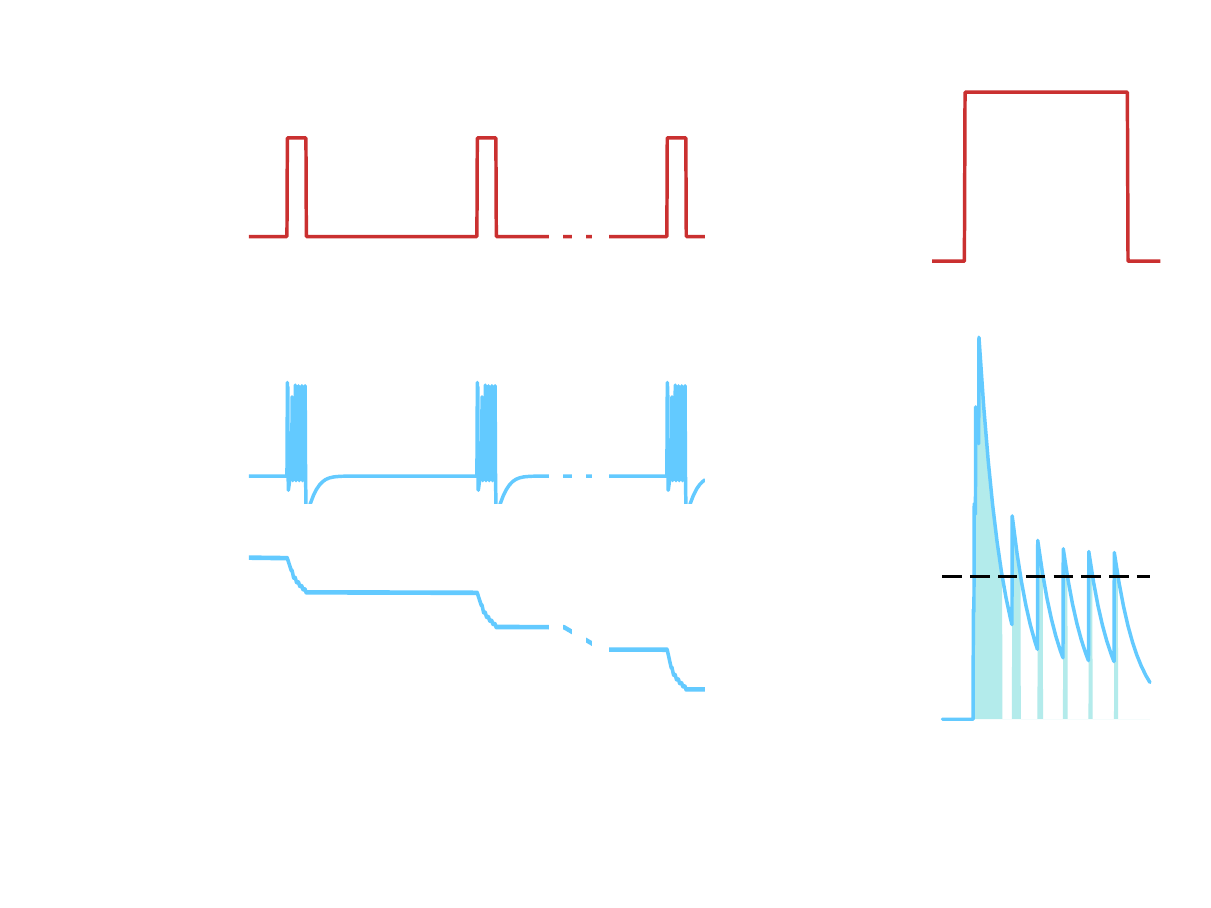
			&
			\resizebox{5cm}{!}{\input{figures2/figure2-b-plasticity.pgf}}
			&
			\resizebox{5cm}{!}{\input{figures2/figure2-steady-power.pgf}}
			
		\end{tabular}
		\\
		\begin{tabular}{p{5cm} p{5cm} p{5cm}}
			\textbf{\large{D}} &  \textbf{\large{E}} & \textbf{\large{F}} \\
			\def\svgwidth{5cm}
			\raisebox{.7\height}{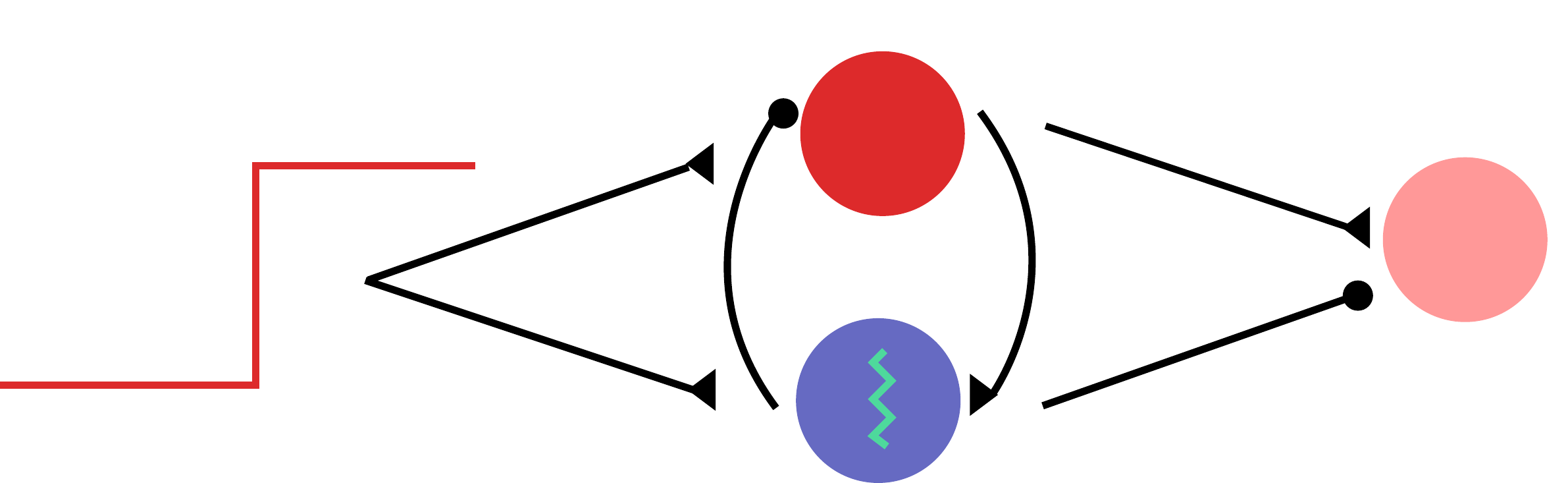}
			&
			\resizebox{5cm}{!}{\input{figures2/figure2-RON-vvm-all-nu-20_step-250_g-2_40_T-100000_A-10_plast-True.pgf}}
			&
			\resizebox{5cm}{!}{\input{figures2/figure2-RON.pgf}}	
			
		\end{tabular}				
		
	\end{tabular}
	\\
	\caption{ {\bf  Model of gap junction plasticity. Bursting induces gLTD, spiking gLTP.} \label{fig2}}
	(\textbf{A}) Bursting protocol replicated from Haas et al. [16]. A current (red line, top panel) of 300 pA for 50 ms at 2 Hz and of -80 pA otherwise injected into a pair of coupled neurons induces repeated bursting (blue line, middle panel, voltage trace). To quantify the amount of bursting, we low-pass filtered ($b_i$) the voltage trace, threshold it at $\theta_{burst} = 1.3$ (discontinued dark line), and integrate. Light blue areas represent the periods during which
bursts are detected and therefore gap junctions are depressed. 
(\textbf{B}) When neurons N1 and N2 spike sparsely (top panel, dark blue, first part of the stimulus), gap junctions are potentiated (bottom panel, green line, first part
of the stimulation), whereas when they are bursting, gap junctions are depressed (second part of the stimulation). 
(\textbf{C}) Green dots show steady-state values of the mean gap junction coupling for the gLTP with soft bounds, for different values of the network drive along the y-axis. With the softbound rule, the steady-state can be found in the AI regime, where the power of the oscillations of the population spiking activity is low (blue area). 
(\textbf{D}) Network architecture: A step excitatory drive is fed to the network of E and I neurons (same network detailed on Figure 1, with plastic gap junction) inducing gamma oscillations. The activity of the network is read out by a downstream population of 200 regular spiking cells.
(\textbf{E}) Top panel, step excitatory drive fed to the networks. Second panel, evolution of the mean gap junction coupling. As the excitatory drive is delivered, a gamma oscillation appears, leading to an increase in bursting activity which is followed by a depression of the gap junctions, until the new fixed point is reached. Bottom panels, raster plots of the inhibitory neurons (blue, I1), excitatory neurons (red, E1) and read-out neurons (red, RON). 6 s of data is represented.
(\textbf{F}) Top panel, step excitatory drive. Other panels, population activity of the read-out neurons in red, evolution of the mean gap junction coupling in light blue. Second panel, simulation with plastic gap junctions. The read-out neurons are the most active during the transient oscillations. Third panel, static gap junction coupling. The read-out neurons are active as long as the excitatory drive is high. Bottom panel, no gap junction coupling. The read-out neurons are not active. 10 s of data is represented.
\end{fullwidth}
\end{figure}

\newpage
%
%

\begin{figure}[H]
\begin{fullwidth}
	\begin{tabular}{ p{14cm}}
	
		\begin{tabular}{p{5cm} p{5cm} p{5.5cm}}
			\textbf{\large{A}} &  \textbf{\large{B}} & \textbf{\large{C}} \\
		 	\resizebox{5cm}{!}{
%
%
%
%
%
%
%
%
%
%
%
%
%
%
\usetikzlibrary{matrix,chains,positioning,decorations.pathreplacing,arrows,arrows.meta}
\usetikzlibrary{matrix,positioning,calc}
\usetikzlibrary{decorations.pathmorphing}

\xdefinecolor{c1}{HTML}{4ED99C}
\xdefinecolor{Ecol}{HTML}{FF6868}
\xdefinecolor{Icol}{HTML}{3366cc}

\xdefinecolor{GN}{HTML}{66ccff}
\xdefinecolor{SN}{HTML}{9999cc}
\xdefinecolor{GAP}{HTML}{4ED99C}

\tikzstyle{Ecircle} = [draw, circle, minimum size=.7cm, fill=Ecol]
\tikzstyle{SNcircle} = [draw, circle, minimum size=.7cm, fill=SN]
\tikzstyle{GNcircle} = [draw, circle, minimum size=.7cm, fill=GN]

\begin{tikzpicture}
\LARGE

\node[Ecircle] (E1) at (0,-1) {E};
\node[GNcircle] (I1) at (0.7,-1.7) {I};
\node[GNcircle] (I2) at (1.3,-2.3) {I};
\node[GNcircle] (I3) at (1.9,-2.9) {I};
\node[GNcircle] (I4) at (2.5,-3.5) {I};

\node [
        draw, rectangle,
        anchor=north west,
        minimum width=3.5cm, minimum height=3.5cm,
        label=Gamma Network
      ] at (-0.5,-0.5) {};

\node [
        draw, rectangle,
        anchor=north west,
        color=GN,
        minimum width=2.7cm, minimum height=2.7cm,
      ] (GN) at (0.3,-1.3) {};


\begin{scope} [color=GAP]
\path
(I1) edge [bend right=60] node {} (I2)
	edge [bend left=60] node {} (I3)
	edge [bend right=60] node {} (I4)
(I2) edge [bend left=60] node {} (I3)
	edge [bend right=60] node {} (I4)
(I3) edge [bend right=60] node {} (I4);
\end{scope}

\node [
        draw, rectangle,
        anchor=north west,
        minimum width=3.5cm, minimum height=3.5cm,
        label=Slow Network
      ] at (3,-4) {};

\node [
        draw, rectangle,
        anchor=north west,
        color=SN,
        minimum width=2.7cm, minimum height=2.7cm,
      ] (SN) at (3,-4) {};

\node[Ecircle] (E2) at (6,-7) {E};
\node[SNcircle] (I5) at (3.5,-4.5) {I};
\node[SNcircle] (I6) at (4.1,-5.1) {I};
\node[SNcircle] (I7) at (4.7,-5.7) {I};
\node[SNcircle] (I8) at (5.3,-6.3) {I};

\begin{scope} [color=GAP]
\path
(I5) edge [bend left=60] node {} (I6)
	edge [bend left=60] node {} (I7)
	edge [bend left=60] node {} (I8)
(I6) edge [bend right=60] node {} (I7)
	edge [bend right=60] node {} (I8)
(I7) edge [bend left=60] node {} (I8);
\end{scope}

\begin{scope}   [dashed,color=GAP] 
\path
	(I3) edge [bend left=60] node {} (I5)
	(I6) edge [bend left=60] node {} (I4)
	(I4) edge node {} (I5);
\end{scope}

\path
	(E1) edge [bend left=50,-triangle 90 reversed] node {} (GN)
	(E1) edge [bend right=50,*-] node {} (GN)
	(E2) edge [bend left=50,-triangle 90 reversed] node {} (SN)
	(E2) edge [bend right=50,*-] node {} (SN);

\end{tikzpicture}
}
			&	
			\resizebox{5cm}{!}{\input{figures3/figure3-resonance.pgf}}
			&
			\resizebox{5cm}{!}{\input{figures3/figure3-mean_vm.pgf}}
			
		\end{tabular}
        \\[-0.2in]
		\begin{tabular}{p{5cm} p{5cm} p{5cm} }
			\textbf{\large{D}} &  \textbf{\large{E}} &  \textbf{\large{F}} \\
			\resizebox{5cm}{!}{
\begingroup%
\makeatletter%
\begin{pgfpicture}%
\pgfpathrectangle{\pgfpointorigin}{\pgfqpoint{5.000000in}{4.000000in}}%
\pgfusepath{use as bounding box, clip}%
\begin{pgfscope}%
\pgfsetbuttcap%
\pgfsetmiterjoin%
\definecolor{currentfill}{rgb}{1.000000,1.000000,1.000000}%
\pgfsetfillcolor{currentfill}%
\pgfsetlinewidth{0.000000pt}%
\definecolor{currentstroke}{rgb}{1.000000,1.000000,1.000000}%
\pgfsetstrokecolor{currentstroke}%
\pgfsetdash{}{0pt}%
\pgfpathmoveto{\pgfqpoint{0.000000in}{0.000000in}}%
\pgfpathlineto{\pgfqpoint{5.000000in}{0.000000in}}%
\pgfpathlineto{\pgfqpoint{5.000000in}{4.000000in}}%
\pgfpathlineto{\pgfqpoint{0.000000in}{4.000000in}}%
\pgfpathclose%
\pgfusepath{fill}%
\end{pgfscope}%
\begin{pgfscope}%
\pgfsetbuttcap%
\pgfsetmiterjoin%
\definecolor{currentfill}{rgb}{1.000000,1.000000,1.000000}%
\pgfsetfillcolor{currentfill}%
\pgfsetlinewidth{0.000000pt}%
\definecolor{currentstroke}{rgb}{0.000000,0.000000,0.000000}%
\pgfsetstrokecolor{currentstroke}%
\pgfsetstrokeopacity{0.000000}%
\pgfsetdash{}{0pt}%
\pgfpathmoveto{\pgfqpoint{0.857224in}{0.842868in}}%
\pgfpathlineto{\pgfqpoint{4.820000in}{0.842868in}}%
\pgfpathlineto{\pgfqpoint{4.820000in}{3.820000in}}%
\pgfpathlineto{\pgfqpoint{0.857224in}{3.820000in}}%
\pgfpathclose%
\pgfusepath{fill}%
\end{pgfscope}%
\begin{pgfscope}%
\definecolor{textcolor}{rgb}{0.150000,0.150000,0.150000}%
\pgfsetstrokecolor{textcolor}%
\pgfsetfillcolor{textcolor}%
\pgftext[x=1.307539in,y=0.745646in,,top]{\color{textcolor}\fontsize{20.000000}{24.000000}\selectfont \(\displaystyle 20\)}%
\end{pgfscope}%
\begin{pgfscope}%
\definecolor{textcolor}{rgb}{0.150000,0.150000,0.150000}%
\pgfsetstrokecolor{textcolor}%
\pgfsetfillcolor{textcolor}%
\pgftext[x=3.108801in,y=0.745646in,,top]{\color{textcolor}\fontsize{20.000000}{24.000000}\selectfont \(\displaystyle 40\)}%
\end{pgfscope}%
\begin{pgfscope}%
\definecolor{textcolor}{rgb}{0.150000,0.150000,0.150000}%
\pgfsetstrokecolor{textcolor}%
\pgfsetfillcolor{textcolor}%
\pgftext[x=2.838612in,y=0.421264in,,top]{\color{textcolor}\fontsize{22.000000}{26.400000}\selectfont Membrane time constant, \(\displaystyle \tau_v\)}%
\end{pgfscope}%
\begin{pgfscope}%
\definecolor{textcolor}{rgb}{0.150000,0.150000,0.150000}%
\pgfsetstrokecolor{textcolor}%
\pgfsetfillcolor{textcolor}%
\pgftext[x=0.505211in,y=1.479576in,left,base]{\color{textcolor}\fontsize{20.000000}{24.000000}\selectfont \(\displaystyle 30\)}%
\end{pgfscope}%
\begin{pgfscope}%
\definecolor{textcolor}{rgb}{0.150000,0.150000,0.150000}%
\pgfsetstrokecolor{textcolor}%
\pgfsetfillcolor{textcolor}%
\pgftext[x=0.505211in,y=2.504527in,left,base]{\color{textcolor}\fontsize{20.000000}{24.000000}\selectfont \(\displaystyle 40\)}%
\end{pgfscope}%
\begin{pgfscope}%
\definecolor{textcolor}{rgb}{0.150000,0.150000,0.150000}%
\pgfsetstrokecolor{textcolor}%
\pgfsetfillcolor{textcolor}%
\pgftext[x=0.505211in,y=3.529478in,left,base]{\color{textcolor}\fontsize{20.000000}{24.000000}\selectfont \(\displaystyle 50\)}%
\end{pgfscope}%
\begin{pgfscope}%
\definecolor{textcolor}{rgb}{0.150000,0.150000,0.150000}%
\pgfsetstrokecolor{textcolor}%
\pgfsetfillcolor{textcolor}%
\pgftext[x=0.449655in,y=2.331434in,,bottom,rotate=90.000000]{\color{textcolor}\fontsize{22.000000}{26.400000}\selectfont Frequency, \(\displaystyle f_{res}\)}%
\end{pgfscope}%
\begin{pgfscope}%
\pgfpathrectangle{\pgfqpoint{0.857224in}{0.842868in}}{\pgfqpoint{3.962776in}{2.977132in}} %
\pgfusepath{clip}%
\pgfsetroundcap%
\pgfsetroundjoin%
\pgfsetlinewidth{2.007500pt}%
\definecolor{currentstroke}{rgb}{0.000000,0.000000,0.000000}%
\pgfsetstrokecolor{currentstroke}%
\pgfsetdash{}{0pt}%
\pgfpathmoveto{\pgfqpoint{1.037350in}{3.365515in}}%
\pgfpathlineto{\pgfqpoint{1.487665in}{2.609869in}}%
\pgfpathlineto{\pgfqpoint{1.937981in}{2.277364in}}%
\pgfpathlineto{\pgfqpoint{2.388296in}{1.971827in}}%
\pgfpathlineto{\pgfqpoint{2.838612in}{1.691071in}}%
\pgfpathlineto{\pgfqpoint{3.288927in}{1.433086in}}%
\pgfpathlineto{\pgfqpoint{3.739243in}{1.196026in}}%
\pgfpathlineto{\pgfqpoint{4.189558in}{0.978192in}}%
\pgfpathlineto{\pgfqpoint{4.639874in}{0.978192in}}%
\pgfusepath{stroke}%
\end{pgfscope}%
\begin{pgfscope}%
\pgfpathrectangle{\pgfqpoint{0.857224in}{0.842868in}}{\pgfqpoint{3.962776in}{2.977132in}} %
\pgfusepath{clip}%
\pgfsetroundcap%
\pgfsetroundjoin%
\pgfsetlinewidth{2.007500pt}%
\definecolor{currentstroke}{rgb}{0.823529,0.831373,0.839216}%
\pgfsetstrokecolor{currentstroke}%
\pgfsetdash{}{0pt}%
\pgfpathmoveto{\pgfqpoint{1.037350in}{3.684676in}}%
\pgfpathlineto{\pgfqpoint{1.487665in}{3.063744in}}%
\pgfpathlineto{\pgfqpoint{1.937981in}{2.442812in}}%
\pgfpathlineto{\pgfqpoint{2.388296in}{2.028858in}}%
\pgfpathlineto{\pgfqpoint{2.838612in}{1.821881in}}%
\pgfpathlineto{\pgfqpoint{3.288927in}{1.407926in}}%
\pgfpathlineto{\pgfqpoint{3.739243in}{1.200949in}}%
\pgfpathlineto{\pgfqpoint{4.189558in}{1.200949in}}%
\pgfpathlineto{\pgfqpoint{4.639874in}{0.993972in}}%
\pgfusepath{stroke}%
\end{pgfscope}%
\begin{pgfscope}%
\pgfsetrectcap%
\pgfsetmiterjoin%
\pgfsetlinewidth{0.000000pt}%
\definecolor{currentstroke}{rgb}{1.000000,1.000000,1.000000}%
\pgfsetstrokecolor{currentstroke}%
\pgfsetdash{}{0pt}%
\pgfpathmoveto{\pgfqpoint{0.857224in}{0.842868in}}%
\pgfpathlineto{\pgfqpoint{0.857224in}{3.820000in}}%
\pgfusepath{}%
\end{pgfscope}%
\begin{pgfscope}%
\pgfsetrectcap%
\pgfsetmiterjoin%
\pgfsetlinewidth{0.000000pt}%
\definecolor{currentstroke}{rgb}{1.000000,1.000000,1.000000}%
\pgfsetstrokecolor{currentstroke}%
\pgfsetdash{}{0pt}%
\pgfpathmoveto{\pgfqpoint{4.820000in}{0.842868in}}%
\pgfpathlineto{\pgfqpoint{4.820000in}{3.820000in}}%
\pgfusepath{}%
\end{pgfscope}%
\begin{pgfscope}%
\pgfsetrectcap%
\pgfsetmiterjoin%
\pgfsetlinewidth{0.000000pt}%
\definecolor{currentstroke}{rgb}{1.000000,1.000000,1.000000}%
\pgfsetstrokecolor{currentstroke}%
\pgfsetdash{}{0pt}%
\pgfpathmoveto{\pgfqpoint{0.857224in}{0.842868in}}%
\pgfpathlineto{\pgfqpoint{4.820000in}{0.842868in}}%
\pgfusepath{}%
\end{pgfscope}%
\begin{pgfscope}%
\pgfsetrectcap%
\pgfsetmiterjoin%
\pgfsetlinewidth{0.000000pt}%
\definecolor{currentstroke}{rgb}{1.000000,1.000000,1.000000}%
\pgfsetstrokecolor{currentstroke}%
\pgfsetdash{}{0pt}%
\pgfpathmoveto{\pgfqpoint{0.857224in}{3.820000in}}%
\pgfpathlineto{\pgfqpoint{4.820000in}{3.820000in}}%
\pgfusepath{}%
\end{pgfscope}%
\begin{pgfscope}%
\pgfsetroundcap%
\pgfsetroundjoin%
\pgfsetlinewidth{2.007500pt}%
\definecolor{currentstroke}{rgb}{0.000000,0.000000,0.000000}%
\pgfsetstrokecolor{currentstroke}%
\pgfsetdash{}{0pt}%
\pgfpathmoveto{\pgfqpoint{1.772563in}{3.492559in}}%
\pgfpathlineto{\pgfqpoint{2.272563in}{3.492559in}}%
\pgfusepath{stroke}%
\end{pgfscope}%
\begin{pgfscope}%
\pgftext[x=2.472563in,y=3.405059in,left,base]{\fontsize{18.000000}{21.600000}\selectfont single neuron reson.}%
\end{pgfscope}%
\begin{pgfscope}%
\pgfsetroundcap%
\pgfsetroundjoin%
\pgfsetlinewidth{2.007500pt}%
\definecolor{currentstroke}{rgb}{0.823529,0.831373,0.839216}%
\pgfsetstrokecolor{currentstroke}%
\pgfsetdash{}{0pt}%
\pgfpathmoveto{\pgfqpoint{1.772563in}{3.122075in}}%
\pgfpathlineto{\pgfqpoint{2.272563in}{3.122075in}}%
\pgfusepath{stroke}%
\end{pgfscope}%
\begin{pgfscope}%
\pgftext[x=2.472563in,y=3.034575in,left,base]{\fontsize{18.000000}{21.600000}\selectfont network frequency}%
\end{pgfscope}%
\end{pgfpicture}%
\makeatother%
\endgroup%
}
			&
			\resizebox{5cm}{!}{\input{figures4/figure4-freq_g1-6_0_g2-6_0_plast_False_cEI-500.pgf}}
            		&
			\resizebox{5cm}{!}{\input{figures4/figure4-power_g1-6_0_g2-6_0_plast_False_cEI-500.pgf}}
	
		\end{tabular}
        \\[-0.2in]
        \begin{tabular}{p{5cm} p{5cm} p{5cm} }
			\textbf{\large{G}} &  \textbf{\large{H}} &  \textbf{\large{I}}  \\
			\resizebox{5cm}{!}{
\begingroup%
\makeatletter%
\begin{pgfpicture}%
\pgfpathrectangle{\pgfpointorigin}{\pgfqpoint{5.000000in}{4.000000in}}%
\pgfusepath{use as bounding box, clip}%
\begin{pgfscope}%
\pgfsetbuttcap%
\pgfsetmiterjoin%
\definecolor{currentfill}{rgb}{1.000000,1.000000,1.000000}%
\pgfsetfillcolor{currentfill}%
\pgfsetlinewidth{0.000000pt}%
\definecolor{currentstroke}{rgb}{1.000000,1.000000,1.000000}%
\pgfsetstrokecolor{currentstroke}%
\pgfsetdash{}{0pt}%
\pgfpathmoveto{\pgfqpoint{0.000000in}{0.000000in}}%
\pgfpathlineto{\pgfqpoint{5.000000in}{0.000000in}}%
\pgfpathlineto{\pgfqpoint{5.000000in}{4.000000in}}%
\pgfpathlineto{\pgfqpoint{0.000000in}{4.000000in}}%
\pgfpathclose%
\pgfusepath{fill}%
\end{pgfscope}%
\begin{pgfscope}%
\pgfsetbuttcap%
\pgfsetmiterjoin%
\definecolor{currentfill}{rgb}{1.000000,1.000000,1.000000}%
\pgfsetfillcolor{currentfill}%
\pgfsetlinewidth{0.000000pt}%
\definecolor{currentstroke}{rgb}{0.000000,0.000000,0.000000}%
\pgfsetstrokecolor{currentstroke}%
\pgfsetstrokeopacity{0.000000}%
\pgfsetdash{}{0pt}%
\pgfpathmoveto{\pgfqpoint{1.045522in}{0.180000in}}%
\pgfpathlineto{\pgfqpoint{4.820000in}{0.180000in}}%
\pgfpathlineto{\pgfqpoint{4.820000in}{3.377686in}}%
\pgfpathlineto{\pgfqpoint{1.045522in}{3.377686in}}%
\pgfpathclose%
\pgfusepath{fill}%
\end{pgfscope}%
\begin{pgfscope}%
\definecolor{textcolor}{rgb}{0.150000,0.150000,0.150000}%
\pgfsetstrokecolor{textcolor}%
\pgfsetfillcolor{textcolor}%
\pgftext[x=0.207469in,y=0.967377in,left,base]{\color{textcolor}\fontsize{20.000000}{24.000000}\selectfont 0 sGJ}%
\end{pgfscope}%
\begin{pgfscope}%
\definecolor{textcolor}{rgb}{0.150000,0.150000,0.150000}%
\pgfsetstrokecolor{textcolor}%
\pgfsetfillcolor{textcolor}%
\pgftext[x=0.030738in,y=2.455136in,left,base]{\color{textcolor}\fontsize{20.000000}{24.000000}\selectfont 40 sGJ}%
\end{pgfscope}%
\begin{pgfscope}%
\pgfsys@transformshift{1.222222in}{1.277778in}%
\pgftext[left,bottom]{\pgfimage[interpolate=true,width=1.569444in,height=0.319444in]{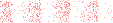}}%
\end{pgfscope}%
\begin{pgfscope}%
\pgfsys@transformshift{1.277778in}{0.986111in}%
\pgftext[left,bottom]{\pgfimage[interpolate=true,width=1.388889in,height=0.305556in]{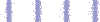}}%
\end{pgfscope}%
\begin{pgfscope}%
\pgfsys@transformshift{1.486111in}{0.611111in}%
\pgftext[left,bottom]{\pgfimage[interpolate=true,width=1.180556in,height=0.305556in]{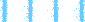}}%
\end{pgfscope}%
\begin{pgfscope}%
\pgfsys@transformshift{1.291667in}{0.319444in}%
\pgftext[left,bottom]{\pgfimage[interpolate=true,width=1.500000in,height=0.305556in]{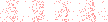}}%
\end{pgfscope}%
\begin{pgfscope}%
\pgfsys@transformshift{1.222222in}{2.916667in}%
\pgftext[left,bottom]{\pgfimage[interpolate=true,width=1.569444in,height=0.305556in]{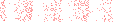}}%
\end{pgfscope}%
\begin{pgfscope}%
\pgfsys@transformshift{1.236111in}{2.625000in}%
\pgftext[left,bottom]{\pgfimage[interpolate=true,width=1.291667in,height=0.305556in]{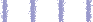}}%
\end{pgfscope}%
\begin{pgfscope}%
\pgfsys@transformshift{1.222222in}{2.250000in}%
\pgftext[left,bottom]{\pgfimage[interpolate=true,width=1.555556in,height=0.305556in]{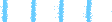}}%
\end{pgfscope}%
\begin{pgfscope}%
\pgfsys@transformshift{1.222222in}{1.958333in}%
\pgftext[left,bottom]{\pgfimage[interpolate=true,width=1.569444in,height=0.305556in]{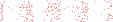}}%
\end{pgfscope}%
\begin{pgfscope}%
\pgfsys@transformshift{3.097222in}{1.277778in}%
\pgftext[left,bottom]{\pgfimage[interpolate=true,width=1.555556in,height=0.319444in]{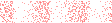}}%
\end{pgfscope}%
\begin{pgfscope}%
\pgfsys@transformshift{3.166667in}{0.986111in}%
\pgftext[left,bottom]{\pgfimage[interpolate=true,width=1.347222in,height=0.305556in]{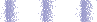}}%
\end{pgfscope}%
\begin{pgfscope}%
\pgfsys@transformshift{3.347222in}{0.611111in}%
\pgftext[left,bottom]{\pgfimage[interpolate=true,width=1.166667in,height=0.305556in]{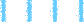}}%
\end{pgfscope}%
\begin{pgfscope}%
\pgfsys@transformshift{3.111111in}{0.319444in}%
\pgftext[left,bottom]{\pgfimage[interpolate=true,width=1.541667in,height=0.305556in]{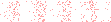}}%
\end{pgfscope}%
\begin{pgfscope}%
\pgfsys@transformshift{3.083333in}{2.916667in}%
\pgftext[left,bottom]{\pgfimage[interpolate=true,width=1.569444in,height=0.305556in]{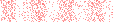}}%
\end{pgfscope}%
\begin{pgfscope}%
\pgfsys@transformshift{3.250000in}{2.625000in}%
\pgftext[left,bottom]{\pgfimage[interpolate=true,width=1.388889in,height=0.305556in]{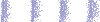}}%
\end{pgfscope}%
\begin{pgfscope}%
\pgfsys@transformshift{3.166667in}{2.250000in}%
\pgftext[left,bottom]{\pgfimage[interpolate=true,width=1.319444in,height=0.305556in]{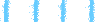}}%
\end{pgfscope}%
\begin{pgfscope}%
\pgfsys@transformshift{3.097222in}{1.958333in}%
\pgftext[left,bottom]{\pgfimage[interpolate=true,width=1.555556in,height=0.305556in]{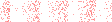}}%
\end{pgfscope}%
\begin{pgfscope}%
\pgfsetrectcap%
\pgfsetmiterjoin%
\pgfsetlinewidth{0.000000pt}%
\definecolor{currentstroke}{rgb}{1.000000,1.000000,1.000000}%
\pgfsetstrokecolor{currentstroke}%
\pgfsetdash{}{0pt}%
\pgfpathmoveto{\pgfqpoint{1.045522in}{0.180000in}}%
\pgfpathlineto{\pgfqpoint{1.045522in}{3.377686in}}%
\pgfusepath{}%
\end{pgfscope}%
\begin{pgfscope}%
\pgfsetrectcap%
\pgfsetmiterjoin%
\pgfsetlinewidth{0.000000pt}%
\definecolor{currentstroke}{rgb}{1.000000,1.000000,1.000000}%
\pgfsetstrokecolor{currentstroke}%
\pgfsetdash{}{0pt}%
\pgfpathmoveto{\pgfqpoint{4.820000in}{0.180000in}}%
\pgfpathlineto{\pgfqpoint{4.820000in}{3.377686in}}%
\pgfusepath{}%
\end{pgfscope}%
\begin{pgfscope}%
\pgfsetrectcap%
\pgfsetmiterjoin%
\pgfsetlinewidth{0.000000pt}%
\definecolor{currentstroke}{rgb}{1.000000,1.000000,1.000000}%
\pgfsetstrokecolor{currentstroke}%
\pgfsetdash{}{0pt}%
\pgfpathmoveto{\pgfqpoint{1.045522in}{0.180000in}}%
\pgfpathlineto{\pgfqpoint{4.820000in}{0.180000in}}%
\pgfusepath{}%
\end{pgfscope}%
\begin{pgfscope}%
\pgfsetrectcap%
\pgfsetmiterjoin%
\pgfsetlinewidth{0.000000pt}%
\definecolor{currentstroke}{rgb}{1.000000,1.000000,1.000000}%
\pgfsetstrokecolor{currentstroke}%
\pgfsetdash{}{0pt}%
\pgfpathmoveto{\pgfqpoint{1.045522in}{3.377686in}}%
\pgfpathlineto{\pgfqpoint{4.820000in}{3.377686in}}%
\pgfusepath{}%
\end{pgfscope}%
\begin{pgfscope}%
\pgftext[x=1.374440in,y=3.304538in,left,base]{\fontsize{18.000000}{21.600000}\selectfont \(\displaystyle \Delta f_{res}: 11\) Hz}%
\end{pgfscope}%
\begin{pgfscope}%
\pgftext[x=3.244252in,y=3.304538in,left,base]{\fontsize{18.000000}{21.600000}\selectfont \(\displaystyle \Delta f_{res}:21\) Hz}%
\end{pgfscope}%
\begin{pgfscope}%
\pgftext[x=2.42761in,y=3.620903in,,base]{\fontsize{22.000000}{26.400000}\selectfont Cross-network synchronization}%
\end{pgfscope}%
\end{pgfpicture}%
\makeatother%
\endgroup%
}
			&
			\resizebox{5cm}{!}{\input{figures4/figure4-phase_g1-6_0_g2-6_0_plast_False_cEI-500.pgf}}
			&
			\resizebox{5cm}{!}{\input{figures4/figure4-cor_g1-6_0_g2-6_0_plast_False_cEI-500.pgf}}
		\end{tabular}
	\end{tabular}
	\\
	\caption{ {\bf Subnetworks having different frequency preferences can synchronize their activity if they share gap junctions.} \label{fig3}}
	(\textbf{A}) Both subnetworks have the same topology with all-to-all connected inhibitory and excitatory neurons. Inhibitory neurons have static gap junctions. The Gamma Network (GN) is connected to the Slow Network (SN) with a varying number of gap junctions. The time constant of the SN inhibitory neuron membranes is varied.
    (\textbf{B}) Frequency-transfer characteristics of one single inhibitory neuron to a subthreshold oscillatory input current (see Methods) for different values of its membrane time constant $\tau_v$. The subthreshold resonance frequency decreases as $\tau_v$ increases.
Data of Cardin et al. 2009 is also represented (black line, figure redrawn from [32] figure 3d).
	(\textbf{C}) Changing the single neuron subthreshold resonance modifies the network oscillation frequency. Mean inhibitory membrane potential for $\tau_v = 17$ ms (continuous line) and $\tau_v = 55$ ms (dashed line). 100 ms of data is represented.
    (\textbf{D}) Relationship between single neuron resonance (black line) and network oscillation frequency (gray line).
    For the following figures E and F, for $(\Delta f_{res}, \# GJs)$, the upper (lower) triangle represents the value in the SN (GN).
    For panels E, F, H, I, the x-axis represents the number of cross-network gap junctions between the GN and SN. The y-axis represents the difference of resonance frequency between the GN and SN. 
    (\textbf{E}) Oscillation frequencies. We observe that the GN and SN adopt the same oscillation frequency for low $\Delta f_{res}$ and high number of shared gap junctions (sGJ).
     (\textbf{F}) Oscillation power. Only increasing $\Delta f_{res}$ seems to have an impact of the power of the SN.
     (\textbf{G}) Raster plots, where dots represent spiking times and each line represent a neuron, for small (first column) and large (second column) differences in $\Delta f_{res}$. For all raster plots, from top to bottom are represented excitatory and inhibitory neurons from SN, then inhibitory and excitatory neurons from GN. 100 neurons are shown for each population. When no gap junctions are shared (bottom row), both networks do not synchronize and are out-of-phase. With 40 shared GJs (top row), the networks synchronize and are in phase for small values of $\Delta f_{res}$. 100 ms of data is represented.
     (\textbf{H}) Phase differences between population activities of the GN and SN, when they share the same frequency. Lighter squares denote parameters for which the phase difference is lower. The GN and SN are considered in phase when the phase difference is zero. Dark blue squares describe a region that is excluded because the GN and SN do not oscillate at the same frequency, therefore cannot be in phase.
     (\textbf{I}) Pearson's correlation of the PAs of the GN and SN. Comparing with panel H, there is high correlation when the GN and SN are in phase.
    \end{fullwidth}
\end{figure}

\newpage
%
%
\begin{figure}[H]
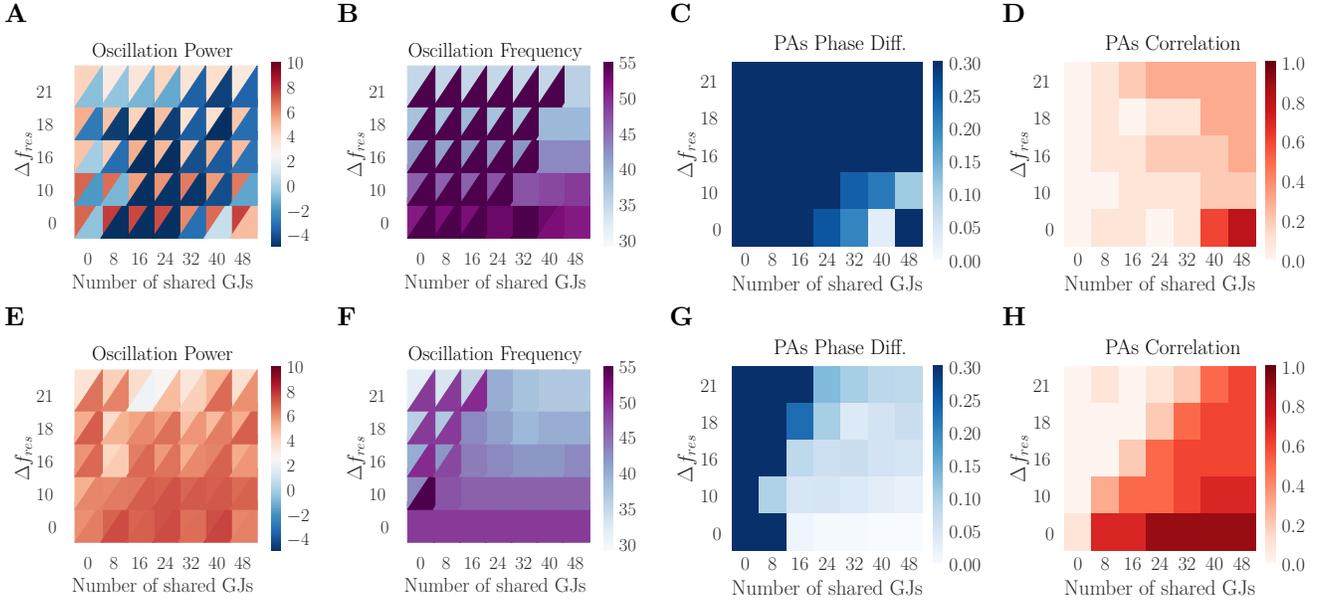

\begin{fullwidth}
	\begin{tabular}{ p{15cm}}

		\begin{tabular}{p{4cm} p{4cm} p{4cm} p{4cm}}
			\textbf{\large{A}} &  \textbf{\large{B}} & \textbf{\large{C}} & \textbf{\large{D}} \\
            		\resizebox{4.5cm}{!}{\input{figures4/figure4-power_g1-6_0_g2-3_0_plast_False_cEI-500.pgf}}
			&
            		\resizebox{4.5cm}{!}{\input{figures4/figure4-freq_g1-6_0_g2-3_0_plast_False_cEI-500.pgf}}
            		&
            		\resizebox{4.5cm}{!}{\input{figures4/figure4-phase_g1-6_0_g2-3_0_plast_False_cEI-500.pgf}}
			&
            		\resizebox{4.5cm}{!}{\input{figures4/figure4-cor_g1-6_0_g2-3_0_plast_False_cEI-500.pgf}}
       	\end{tabular}
		
		\begin{tabular}{p{4cm} p{4cm} p{4cm} p{4cm}}
			\textbf{\large{E}} & \textbf{\large{F}} & \textbf{\large{G}} & \textbf{\large{H}}  \\

            		\resizebox{4.5cm}{!}{\input{figures4/figure4-power_g1-6_0_g2-3_0_plast_True_cEI-500.pgf}}
			&

            		\resizebox{4.5cm}{!}{\input{figures4/figure4-freq_g1-6_0_g2-3_0_plast_True_cEI-500.pgf}}
            		&

            		\resizebox{4.5cm}{!}{\input{figures4/figure4-phase_g1-6_0_g2-3_0_plast_True_cEI-500.pgf}}
			&

            		\resizebox{4.5cm}{!}{\input{figures4/figure4-cor_g1-6_0_g2-3_0_plast_True_cEI-500.pgf}}
		\end{tabular}

	\end{tabular}
	\\
	\caption{ {\bf  Gap junction plasticity lets networks recover synchronization.} \label{fig4}}
	For all panels, the x-axis represents the number of cross-network gap junctions between GN and SN. The y-axis represents the difference of resonance frequency between GN and SN. The gap junctions are static from panels A to D and plastic from panels E to H. Values for the Gamma Network (resp. Slow Network) are represented by the lower (upper) triangles. The GN (SN) has weak (strong) initial mean GJ coupling. Shared GJs are initialized with mean coupling strength in the middle between those of the GN and SN.
    (\textbf{A}) Oscillation power. The GN, with weak GJ coupling, shows weak oscillations.
    (\textbf{B}) Oscillation frequency. We observe that the GN and SN oscillate at the same frequency only for high number of shared GJs.
    (\textbf{C}) Phase differences between PAs of the GN and SN (as for Figure 3H). The GN and SN stay mostly out-of-phase.
 	(\textbf{D}) Correlation of the PAs of the GN and SN. Except for the particular case where $\Delta f_{res}=0$ and the number of shared GJs is high, the PAs of the GN and SN show no correlation.
    (\textbf{E}) Oscillation power. Comparing with panel A, we observe that the oscillation power seems to match in both networks, with mostly the oscillation power of GN (initially weak) increasing to the SN's levels (initially strong).
    (\textbf{F}) Oscillation frequency. Comparing with panel B, we observe an extension of the region where the GN and SN oscillate at the same frequency.
    (\textbf{G}) Phase differences between PAs of the GN and SN. We observe here a large region where the GN and SN are in-phase.
 	(\textbf{H}) Correlation of the PAs of the GN and SN. Comparing with panel D, we observe a large extension of the region where both networks are synchronized. 
    
    \end{fullwidth}
\end{figure}

%
%
\newpage
\begin{figure}[H]
\begin{fullwidth}
	\begin{tabular}{ p{16cm}}
		\begin{tabular}{p{4cm} p{4cm} p{4cm} p{4cm}}
        		\textbf{\large{A}} & \textbf{\large{B}} & \textbf{\large{C}} & \textbf{\large{D}} \\
			\resizebox{4cm}{!}{\input{figures5/figure5-v_traces.pgf}}
            		&
			\resizebox{4cm}{!}{\input{figures5/figure5-non-averaged.pgf}}
            		&
			\resizebox{4cm}{!}{\input{figures5/figure5-histograms.pgf}}
             		&
             		
			\resizebox{4cm}{!}{
\begingroup%
\makeatletter%
\begin{pgfpicture}%
\pgfpathrectangle{\pgfpointorigin}{\pgfqpoint{5.000000in}{4.000000in}}%
\pgfusepath{use as bounding box, clip}%
\begin{pgfscope}%
\pgfsetbuttcap%
\pgfsetmiterjoin%
\definecolor{currentfill}{rgb}{1.000000,1.000000,1.000000}%
\pgfsetfillcolor{currentfill}%
\pgfsetlinewidth{0.100375pt}%
\definecolor{currentstroke}{rgb}{1.000000,1.000000,1.000000}%
\pgfsetstrokecolor{currentstroke}%
\pgfsetdash{}{0pt}%
\pgfpathmoveto{\pgfqpoint{0.000000in}{0.000000in}}%
\pgfpathlineto{\pgfqpoint{5.000000in}{0.000000in}}%
\pgfpathlineto{\pgfqpoint{5.000000in}{4.000000in}}%
\pgfpathlineto{\pgfqpoint{0.000000in}{4.000000in}}%
\pgfpathclose%
\pgfusepath{stroke,fill}%
\end{pgfscope}%
\begin{pgfscope}%
\pgfsetbuttcap%
\pgfsetmiterjoin%
\definecolor{currentfill}{rgb}{1.000000,1.000000,1.000000}%
\pgfsetfillcolor{currentfill}%
\pgfsetlinewidth{0.000000pt}%
\definecolor{currentstroke}{rgb}{0.000000,0.000000,0.000000}%
\pgfsetstrokecolor{currentstroke}%
\pgfsetstrokeopacity{0.000000}%
\pgfsetdash{}{0pt}%
\pgfpathmoveto{\pgfqpoint{1.107680in}{0.879554in}}%
\pgfpathlineto{\pgfqpoint{4.039352in}{0.879554in}}%
\pgfpathlineto{\pgfqpoint{4.039352in}{3.792310in}}%
\pgfpathlineto{\pgfqpoint{1.107680in}{3.792310in}}%
\pgfpathclose%
\pgfusepath{fill}%
\end{pgfscope}%
\begin{pgfscope}%
\definecolor{textcolor}{rgb}{0.150000,0.150000,0.150000}%
\pgfsetstrokecolor{textcolor}%
\pgfsetfillcolor{textcolor}%
\pgftext[x=1.336717in,y=0.782332in,,top]{\color{textcolor}\fontsize{20.000000}{24.000000}\selectfont \(\displaystyle 12.5\)}%
\end{pgfscope}%
\begin{pgfscope}%
\definecolor{textcolor}{rgb}{0.150000,0.150000,0.150000}%
\pgfsetstrokecolor{textcolor}%
\pgfsetfillcolor{textcolor}%
\pgftext[x=2.481901in,y=0.782332in,,top]{\color{textcolor}\fontsize{20.000000}{24.000000}\selectfont \(\displaystyle 15.0\)}%
\end{pgfscope}%
\begin{pgfscope}%
\definecolor{textcolor}{rgb}{0.150000,0.150000,0.150000}%
\pgfsetstrokecolor{textcolor}%
\pgfsetfillcolor{textcolor}%
\pgftext[x=3.627086in,y=0.782332in,,top]{\color{textcolor}\fontsize{20.000000}{24.000000}\selectfont \(\displaystyle 17.5\)}%
\end{pgfscope}%
\begin{pgfscope}%
\definecolor{textcolor}{rgb}{0.150000,0.150000,0.150000}%
\pgfsetstrokecolor{textcolor}%
\pgfsetfillcolor{textcolor}%
\pgftext[x=2.573516in,y=0.457950in,,top]{\color{textcolor}\fontsize{22.000000}{26.400000}\selectfont Peak intervals, \(\displaystyle T_i\) [ms]}%
\end{pgfscope}%
\begin{pgfscope}%
\definecolor{textcolor}{rgb}{0.150000,0.150000,0.150000}%
\pgfsetstrokecolor{textcolor}%
\pgfsetfillcolor{textcolor}%
\pgftext[x=0.883063in,y=0.906439in,left,base]{\color{textcolor}\fontsize{20.000000}{24.000000}\selectfont \(\displaystyle 0\)}%
\end{pgfscope}%
\begin{pgfscope}%
\definecolor{textcolor}{rgb}{0.150000,0.150000,0.150000}%
\pgfsetstrokecolor{textcolor}%
\pgfsetfillcolor{textcolor}%
\pgftext[x=0.628272in,y=1.585408in,left,base]{\color{textcolor}\fontsize{20.000000}{24.000000}\selectfont \(\displaystyle 500\)}%
\end{pgfscope}%
\begin{pgfscope}%
\definecolor{textcolor}{rgb}{0.150000,0.150000,0.150000}%
\pgfsetstrokecolor{textcolor}%
\pgfsetfillcolor{textcolor}%
\pgftext[x=0.500877in,y=2.264377in,left,base]{\color{textcolor}\fontsize{20.000000}{24.000000}\selectfont \(\displaystyle 1000\)}%
\end{pgfscope}%
\begin{pgfscope}%
\definecolor{textcolor}{rgb}{0.150000,0.150000,0.150000}%
\pgfsetstrokecolor{textcolor}%
\pgfsetfillcolor{textcolor}%
\pgftext[x=0.500877in,y=2.943345in,left,base]{\color{textcolor}\fontsize{20.000000}{24.000000}\selectfont \(\displaystyle 1500\)}%
\end{pgfscope}%
\begin{pgfscope}%
\definecolor{textcolor}{rgb}{0.150000,0.150000,0.150000}%
\pgfsetstrokecolor{textcolor}%
\pgfsetfillcolor{textcolor}%
\pgftext[x=0.500877in,y=3.622314in,left,base]{\color{textcolor}\fontsize{20.000000}{24.000000}\selectfont \(\displaystyle 2000\)}%
\end{pgfscope}%
\begin{pgfscope}%
\definecolor{textcolor}{rgb}{0.150000,0.150000,0.150000}%
\pgfsetstrokecolor{textcolor}%
\pgfsetfillcolor{textcolor}%
\pgftext[x=0.445321in,y=2.335932in,,bottom,rotate=90.000000]{\color{textcolor}\fontsize{22.000000}{26.400000}\selectfont Input Amplitudes, \(\displaystyle \overline{x}_i\)}%
\end{pgfscope}%
\begin{pgfscope}%
\pgfsys@transformshift{1.111111in}{0.986111in}%
\pgftext[left,bottom]{\pgfimage[interpolate=true,width=2.930556in,height=2.708333in]{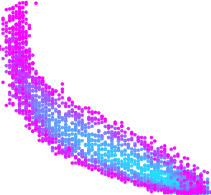}}%
\end{pgfscope}%
\begin{pgfscope}%
\pgfsetrectcap%
\pgfsetmiterjoin%
\pgfsetlinewidth{0.000000pt}%
\definecolor{currentstroke}{rgb}{1.000000,1.000000,1.000000}%
\pgfsetstrokecolor{currentstroke}%
\pgfsetdash{}{0pt}%
\pgfpathmoveto{\pgfqpoint{1.107680in}{0.879554in}}%
\pgfpathlineto{\pgfqpoint{1.107680in}{3.792310in}}%
\pgfusepath{}%
\end{pgfscope}%
\begin{pgfscope}%
\pgfsetrectcap%
\pgfsetmiterjoin%
\pgfsetlinewidth{0.000000pt}%
\definecolor{currentstroke}{rgb}{1.000000,1.000000,1.000000}%
\pgfsetstrokecolor{currentstroke}%
\pgfsetdash{}{0pt}%
\pgfpathmoveto{\pgfqpoint{4.039352in}{0.879554in}}%
\pgfpathlineto{\pgfqpoint{4.039352in}{3.792310in}}%
\pgfusepath{}%
\end{pgfscope}%
\begin{pgfscope}%
\pgfsetrectcap%
\pgfsetmiterjoin%
\pgfsetlinewidth{0.000000pt}%
\definecolor{currentstroke}{rgb}{1.000000,1.000000,1.000000}%
\pgfsetstrokecolor{currentstroke}%
\pgfsetdash{}{0pt}%
\pgfpathmoveto{\pgfqpoint{1.107680in}{0.879554in}}%
\pgfpathlineto{\pgfqpoint{4.039352in}{0.879554in}}%
\pgfusepath{}%
\end{pgfscope}%
\begin{pgfscope}%
\pgfsetrectcap%
\pgfsetmiterjoin%
\pgfsetlinewidth{0.000000pt}%
\definecolor{currentstroke}{rgb}{1.000000,1.000000,1.000000}%
\pgfsetstrokecolor{currentstroke}%
\pgfsetdash{}{0pt}%
\pgfpathmoveto{\pgfqpoint{1.107680in}{3.792310in}}%
\pgfpathlineto{\pgfqpoint{4.039352in}{3.792310in}}%
\pgfusepath{}%
\end{pgfscope}%
\begin{pgfscope}%
\pgfpathrectangle{\pgfqpoint{4.222581in}{0.879554in}}{\pgfqpoint{0.145638in}{2.912756in}} %
\pgfusepath{clip}%
\pgfsetbuttcap%
\pgfsetmiterjoin%
\definecolor{currentfill}{rgb}{1.000000,1.000000,1.000000}%
\pgfsetfillcolor{currentfill}%
\pgfsetlinewidth{0.010037pt}%
\definecolor{currentstroke}{rgb}{1.000000,1.000000,1.000000}%
\pgfsetstrokecolor{currentstroke}%
\pgfsetdash{}{0pt}%
\pgfpathmoveto{\pgfqpoint{4.222581in}{0.879554in}}%
\pgfpathlineto{\pgfqpoint{4.222581in}{0.890932in}}%
\pgfpathlineto{\pgfqpoint{4.222581in}{3.780932in}}%
\pgfpathlineto{\pgfqpoint{4.222581in}{3.792310in}}%
\pgfpathlineto{\pgfqpoint{4.368219in}{3.792310in}}%
\pgfpathlineto{\pgfqpoint{4.368219in}{3.780932in}}%
\pgfpathlineto{\pgfqpoint{4.368219in}{0.890932in}}%
\pgfpathlineto{\pgfqpoint{4.368219in}{0.879554in}}%
\pgfpathclose%
\pgfusepath{stroke,fill}%
\end{pgfscope}%
\begin{pgfscope}%
\definecolor{textcolor}{rgb}{0.150000,0.150000,0.150000}%
\pgfsetstrokecolor{textcolor}%
\pgfsetfillcolor{textcolor}%
\pgftext[x=4.465442in,y=0.774031in,left,base]{\color{textcolor}\fontsize{20.000000}{24.000000}\selectfont \(\displaystyle 0\)}%
\end{pgfscope}%
\begin{pgfscope}%
\definecolor{textcolor}{rgb}{0.150000,0.150000,0.150000}%
\pgfsetstrokecolor{textcolor}%
\pgfsetfillcolor{textcolor}%
\pgftext[x=4.465442in,y=1.371520in,left,base]{\color{textcolor}\fontsize{20.000000}{24.000000}\selectfont \(\displaystyle 200\)}%
\end{pgfscope}%
\begin{pgfscope}%
\definecolor{textcolor}{rgb}{0.150000,0.150000,0.150000}%
\pgfsetstrokecolor{textcolor}%
\pgfsetfillcolor{textcolor}%
\pgftext[x=4.465442in,y=1.969008in,left,base]{\color{textcolor}\fontsize{20.000000}{24.000000}\selectfont \(\displaystyle 400\)}%
\end{pgfscope}%
\begin{pgfscope}%
\definecolor{textcolor}{rgb}{0.150000,0.150000,0.150000}%
\pgfsetstrokecolor{textcolor}%
\pgfsetfillcolor{textcolor}%
\pgftext[x=4.465442in,y=2.566497in,left,base]{\color{textcolor}\fontsize{20.000000}{24.000000}\selectfont \(\displaystyle 600\)}%
\end{pgfscope}%
\begin{pgfscope}%
\definecolor{textcolor}{rgb}{0.150000,0.150000,0.150000}%
\pgfsetstrokecolor{textcolor}%
\pgfsetfillcolor{textcolor}%
\pgftext[x=4.465442in,y=3.163985in,left,base]{\color{textcolor}\fontsize{20.000000}{24.000000}\selectfont \(\displaystyle 800\)}%
\end{pgfscope}%
\begin{pgfscope}%
\pgfsys@transformshift{4.222222in}{0.875000in}%
\pgftext[left,bottom]{\pgfimage[interpolate=true,width=0.152778in,height=2.916667in]{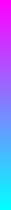}}%
\end{pgfscope}%
\begin{pgfscope}%
\pgfsetbuttcap%
\pgfsetmiterjoin%
\pgfsetlinewidth{0.000000pt}%
\definecolor{currentstroke}{rgb}{1.000000,1.000000,1.000000}%
\pgfsetstrokecolor{currentstroke}%
\pgfsetdash{}{0pt}%
\pgfpathmoveto{\pgfqpoint{4.222581in}{0.879554in}}%
\pgfpathlineto{\pgfqpoint{4.222581in}{0.890932in}}%
\pgfpathlineto{\pgfqpoint{4.222581in}{3.780932in}}%
\pgfpathlineto{\pgfqpoint{4.222581in}{3.792310in}}%
\pgfpathlineto{\pgfqpoint{4.368219in}{3.792310in}}%
\pgfpathlineto{\pgfqpoint{4.368219in}{3.780932in}}%
\pgfpathlineto{\pgfqpoint{4.368219in}{0.890932in}}%
\pgfpathlineto{\pgfqpoint{4.368219in}{0.879554in}}%
\pgfpathclose%
\pgfusepath{}%
\end{pgfscope}%
\end{pgfpicture}%
\makeatother%
\endgroup%
}

		\end{tabular}
        
		\begin{tabular}{p{4cm} p{4cm} p{4cm} p{4cm}}
			\textbf{\large{E}} & \textbf{\large{F}} & \textbf{\large{G}} & \textbf{\large{H}}  \\
            	
			\resizebox{4cm}{!}{\input{figures5/figure5-period-decoded-4hz.pgf}}
            &
			\resizebox{4cm}{!}{\input{figures5/figure5-bandwidth.pgf}}
			&
			\resizebox{4cm}{!}{\input{figures5/figure5-bandwidth-amplitude.pgf}}
			&
			\resizebox{4cm}{!}{\input{figures5/figure5-decoded-input-example.pgf}}
				
		\end{tabular}
        							
        \begin{tabular}{p{8cm} p{4cm} p{4cm}}
			\textbf{\large{I}} & \textbf{\large{J}} & \textbf{\large{K}} \\
			\resizebox{8cm}{!}{\input{figures5/figure5-corr-decoded.pgf}}
            &
			\resizebox{4cm}{!}{\input{figures5/figure5-ctc-robustness.pgf}}
			&
			\resizebox{4cm}{!}{\input{figures5/figure5-ctc-robustness-gap.pgf}}
	
		\end{tabular}

	\end{tabular}
	\\
	\caption{ {\bf Gap junction coupling allows networks to transmit information and gap junction plasticity improves robustness of the transfer.} \label{fig5}}
	(\textbf{A} Voltages traces of inhibitory neurons in the input-network (IN) in light blue and in the output-network (ON) in purple, when networks share no GJs (first rows) or 40 GJs (bottom rows). Despite not directly receiving the input signal, the ON synchronizes its activity with IN.
    For panels B to I, the networks share 40 GJs. 50 ms of data is represented. For the following figures 5B, 5C and 5H, 1 s of data is represented.
    (\textbf{B}) Input signal in red, number of spiking events of inhibitory neurons of IN in light blue and of ON in purple, for time bins of 0.1 ms. 
    (\textbf{C}) Input signal in red, number of spiking events of inhibitory neurons of IN in light blue and ON in purple, for time bins of 25 ms. 
    (\textbf{D}) Input signal amplitude $A_i$ as function of the corresponding PA peak interval $T_i$ for input signals oscillation at 4 Hz with mean varying from 0 to 1000 (See Methods).
    (\textbf{E}) Input signal in red and decoded input signal in purple. The PA peak interval $T_i$ is used to estimate the input amplitude. 
    (\textbf{F}) Correlation between input signal and decoded input signal. The amplitude of the input is 400 pA, its frequency goes from 0 to 100 Hz.
    (\textbf{G}) Correlation between input signal and decoded input signal. The amplitude of the input goes from 0 to 10000 pA, its frequency goes from 0 to 100 Hz.
    (\textbf{H}) Example of 1 s of colored noise input signal (A = 800 pA, mean = 400 pA, $\tau_{filter} = 100$ ms) in red and decoded input in purple (correlation 0.8). 
    (\textbf{I}) Pearson's correlation coefficient between input and decoded input for static (plastic) network in black (gray) for different values of the mean initial GJ coupling strength, as function of the number of shared GJs. The simulation is repeated for 5 different inputs. 
    (\textbf{J}) Pearson's correlation coefficient between input and decoded input for static (resp. plastic) network in black (resp. gray) as function of the proportion of GJs removed. The simulation is repeated for 10 different inputs.
    (\textbf{K}) Mean gap junction change between the steady-state value obtained with all the gap junctions, and the steady-state value obtained after gap junction removal. The remaining gap junctions compensate for the missing ones as they become stronger in strength.
    \end{fullwidth}
\end{figure}

\end{document}